# Scalable Composites Benefiting from Transition Metal Oxides as Cathode Material for Efficient Lithium-Sulfur Battery

Dr. Vittorio Marangon,[a] Eugenio Scaduti,[b] Viviana Fatima Vinci,[b] Prof. Jusef Hassoun[a,b,c,]*

[a] *Graphene Labs, Istituto Italiano di Tecnologia, via Morego 30, Genova, 16163, Italy*

[b] *Department of Chemical, Pharmaceutical and Agricultural Sciences, University of Ferrara, Via Fossato di Mortara 17, Ferrara, 44121, Italy*

[c] *National Interuniversity Consortium of Materials Science and Technology (INSTM), University of Ferrara Research Unit, Via Fossato di Mortara, 17, 44121, Ferrara, Italy.*

*Corresponding author. E-mail addresses: jusef.hassoun@unife.it, jusef.hassoun@iit.it.

## Abstract

Composites materials achieved by including transition metal oxides with different structure and morphology in sulfur are suggested as scalable cathodes for high-energy lithium-sulfur (Li-S) battery. The composites contain 80 wt.% of sulfur and 20 wt.% of either $MnO_2$ or $TiO_2$ that lead to a sulfur content in the electrode of 64 wt.%, and reveal in cell a reversible, fast and lowly polarized conversion process with limited interphase resistance. The S-$TiO_2$ composite exhibits excellent rate capability between C/10 and 2C, and a cycle life extended over 400 cycles at 2C due to the effects of the nanometric $TiO_2$ additive in boosting the reaction kinetics. Instead, the micrometric sized particles of $MnO_2$ partially limit the electrochemical activity of S-$MnO_2$ to the current rate of 1C. Nevertheless, both S-$MnO_2$ and S-$TiO_2$ stand a sulfur loading increased up to values approaching 6 mg cm$^{-2}$, and deliver at C/5 an areal capacity ranging from about 4.5 to 5.5 mAh cm$^{-2}$. The excellent performances of the metal oxide-sulfur electrodes, even at high active material loading, and the possible scalability of the synthetic pathway adopted in the work actually suggest the composites as viable cathodes for next generation Li-S battery with high energy density and efficient electrochemical process.

## Table of Content

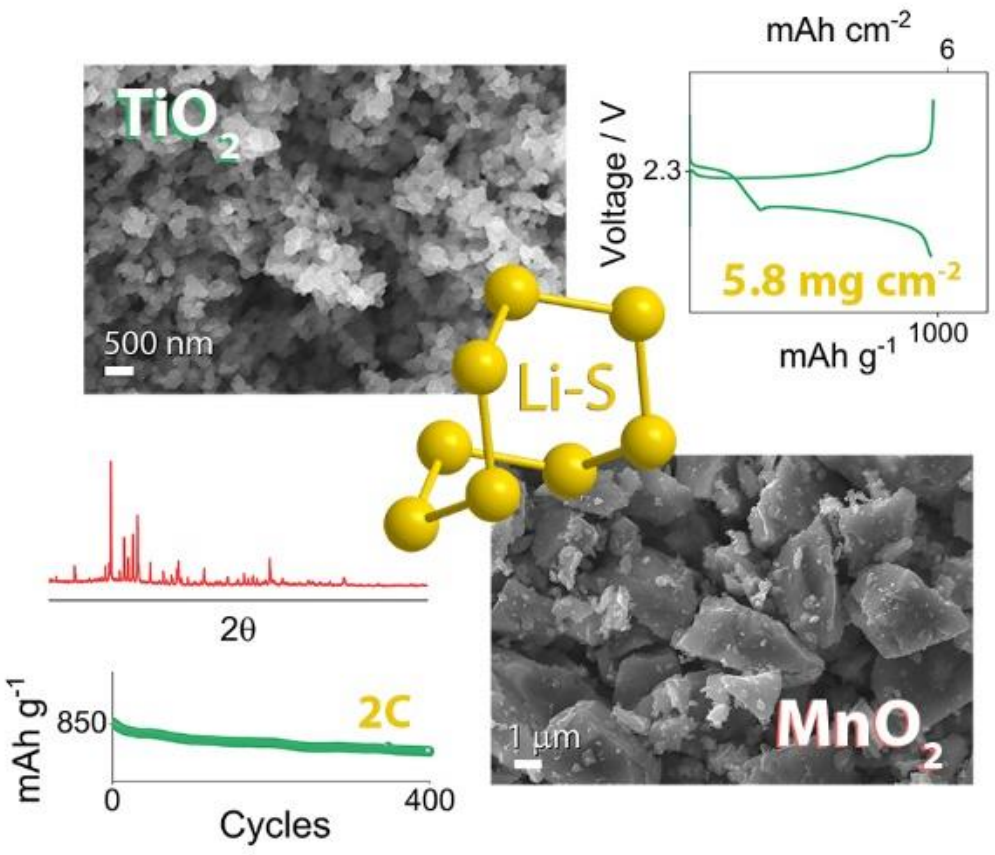

**Scalable Li-S batteries** benefit from sulfur cathodes including either $MnO_2$ or $TiO_2$ as additive to achieve practical energy storage systems with high-capacity and extended cycle life.

## Introduction

Recent concerns raised by the global climate change triggered the development of new energy storage technologies to actually promote the large-scale diffusion of sustainable transport systems, such as electric vehicles (EVs), with satisfactory autonomy.[1–3] The present technology of choice to power most of the electronic devices is the rechargeable lithium-ion battery, that relies on the reversible (de)intercalation of $Li^+$ ions into and from electrodes typically with a layered structure.[4] Despite the notable progress achieved by intensive research work, challenges in the Li-ion technology such as the relatively limited energy content, the environmental and economic impact of the materials, and the need for an efficient recycling process still hinder a complete transition to a zero-emission mobility.[5] In this scenario, rechargeable lithium-sulfur (Li-S) battery may represent an alternative system due the low cost of sulfur, and the high theoretical energy associated with the multi-electron electrochemical conversion reaction summarized by equation (1):[6]

$$16Li^+ + S_8 + 16e^- \rightleftarrows 8Li_2S \quad (1)$$

This intriguing process can in fact deliver an energy density of 3730 Wh kg$^{-1}$ as referred to the sulfur mass, that may lead to a practical value exceeding 400 Wh kg$^{-1}$, thus remarkably improving the one ascribed to the Li-ion which typically ranges from 250 Wh kg$^{-1}$ to about 300 Wh kg$^{-1}$.[6–8] These appealing features have boosted relevant research efforts towards the Li-S technology for addressing the several issues that still prevent its practical and large-scale diffusion.[9–11] Among the various drawbacks affecting the Li-S battery, the most severe may be ascribed with the possible reactivity of the polysulfide intermediates formed during Li-S conversion process ($Li_2S_x$, $2 \leq x \leq 8$) with the anode. Indeed, the soluble fraction of these species formed during discharge (typically $Li_2S_8$ and $Li_2S_6$) can migrate through the electrolyte to undergo direct reduction at the lithium metal surface, and subsequently diffuse back to the cathode to be newly oxidized during charge. This "shuttle process" between the electrodes typically leads to an apparent charge without any energy storage, with active material loss, anode degradation, low coulombic efficiency, and capacity fading.[12,13] Several approaches have been undertaken to control the undesired processes ascribed to the soluble polysulfides during battery operation, including the use of sulfur-carbon composites,[14–17] inorganic cathode additives,[18–20] alternative binders,[21,22] polymer electrolytes,[23,24] electrolyte sacrificial agents,[25–27] as well as functional interlayers[28,29] and separators.[30,31] In particular, transition metal oxides of various nature and morphology have been thoroughly studied for possibly mitigating the polysulfide diffusion into the electrolyte, and improving of the Li-S cell behavior. The favorable effects of metal oxide additives have been attributed to the strong polar interactions between the electron attractive metal nuclei with the negatively charged sulfur ($S_x^{-2}$), and of the oxygens with the $Li^+$ ions.[32–35] Among the various oxides, those based on Mn and Ti have received particular attention from the scientific community due to their exceptional polysulfides restraining ability.[32–35] Furthermore, the development of sustainable sulfur composite cathodes,[36–38] and the implementation of scalable synthesis procedures appeared as very promising approaches for limiting the economic impact and actually achieving a practical Li-S battery.[39] Promising recent works revealed remarkable Li-S performances based on metal oxides having particular morphologies obtained through alternative synthesis pathways, such as three-dimensional porous $MnO_2$[40] and room-like $TiO_2$[41] as hosts for sulfur. In this work, we use the sustainable and environmentally friendly sulfur with transition metal oxides to achieve high-performance composites for lithium battery, through a simple and straightforward physical mixing pathway. Hence, commercial $MnO_2$ and $TiO_2$ powders are employed without additional synthetic steps as advantageous additives in sulfur composites with a weight ratio of the active material as high as 80 %. Electrochemical tests are preliminarily carried out on the $MnO_2$ and $TiO_2$ materials to exclude possible side activity of the oxides which may affect the Li-S conversion process. Subsequently, the corresponding sulfur composites cast on porous carbon collector of suitable electrical conductivity are employed as the positive electrode in efficient lithium cells.[42,43] Hence, we propose herein a possible strategy to achieve improved electrodes, and possibly scale up a high-energy Li-S battery.

## Results and discussion

Structure, morphology and possible side electrochemical activity of the metal oxide additive may actually play a crucial role in determining the cell performance of the corresponding sulfur composites.[32–35] Therefore, $MnO_2$ and $TiO_2$ powders are initially investigated by X-ray diffraction (XRD) as shown in Figure S1a,b (Supporting Information). The corresponding patterns reveal a well-defined crystalline structure for both the metal oxides, identified by $\beta$-phase (pyrolusite) for $MnO_2$ (Fig. S1a) with the $P4_2/mnm$ space group, and $\alpha$ phase (anatase) for $TiO_2$ **(**Fig. S1b**)** with the $I4_1/amdZ$ space group.[44,45] The absence of significant impurities in the XRD patterns suggest the suitability of the oxides for electrochemical application. The morphological features of $MnO_2$ and $TiO_2$ are provided by the scanning electron microscopy (SEM) images reported in Fig. S1(c, e and g) for $MnO_2$ and in Fig. S1(d, f and h) for $TiO_2$. The metal oxide powders exhibit substantial differences in morphology: indeed, $MnO_2$ shows particles with defined flake-like geometry and irregular sizes ranging from sub-micrometric values to about 3 µm (Fig. S1c, e and g), instead $TiO_2$ powder is homogeneously formed by nanometric crystals smaller than 500 nm (Fig. S1d, f and h). It is worth mentioning that large $MnO_2$ particles can possibly hinder the electrolyte decomposition in the lithium cell, while the smaller domains typically boost the kinetics of the electrochemical reactions and favor the discharge and charge processes.[46] On the other hand, the nanometric size observed for the $TiO_2$ particles are particularly suggested to shorten the diffusion path and improve the kinetics of the electrochemical processes. However, excessively small particles typically induce electrolyte decomposition and raise the electrode/electrolyte interphase resistance in particular at increased anodic potentials.[47] A more detailed study on the particle size of $MnO_2$ and $TiO_2$ is provided by the transmission electron microscopy (TEM) images displayed in panels (a, c) and (b, d) of Figure S2 (Supporting Information), respectively, which were acquired upon samples sonication (see experimental section). The analyses confirm the vast size range of the $MnO_2$ flakes, extending from 50 nm to 3 µm, and show wide $TiO_2$ clusters composed of quasi-spherical primary particles with dimensions from 50 to 200 nm. The electrochemical characteristics of $MnO_2$ and $TiO_2$ in lithium cell are then investigated by cyclic voltammetry (CV), galvanostatic cycling and electrochemical impedance spectroscopy (EIS) with outcomes displayed in Figure 1. The voltammogram related to the $MnO_2$ electrode (Fig. 1a) shows reduction with a peak centered at about 2.7 V *vs.* $Li^+$/Li during the cathodic scan, and oxidation at about 3.2 V *vs.* $Li^+$/Li during the reverse cathodic scan as likely due to the electrochemical process reported in equation (2):[48]

$$xLi^+ + MnO_2 + xe^- \rightleftarrows Li_xMnO_2 \quad (2)$$ [49]

The subsequent CV curves become mor symmetric, overlap and indicate a redox process centered at about 3 V *vs.* $Li^+$/Li, while the anodic current slightly improves. The observed changes typically suggest a structural reorganization of the electrode upon the electrochemical process, and the concomitant formation of a solid electrolyte interphase (SEI) layer at the electrodes surface, which may affect the $Li^+$ exchange.[50,51] It is worth mentioning that the relatively low current value for the peaks observed in the voltammogram of Fig. 1a, i.e., with a maximum value approaching 60 µA using an electrode with geometric surface of 0.785 $cm^2$, suggests a poor electrochemical activity of the bare $MnO_2$. On the other hand, $TiO_2$ exhibits CV signals (Fig. 1b) with cathodic peak centered at about 1.7 V *vs.* $Li^+$/Li, and a subsequent anodic one at about 2.1 V *vs.* $Li^+$/Li which identify the (de)insertion process displayed in equation (3):[52]

$$xLi^+ + TiO_2 + xe^- \rightleftarrows Li_xTiO_2 \quad (3)$$

Despite the higher currents for $TiO_2$ compared to $MnO_2$ during the first CV cycle, with anodic and cathodic values of 200 and -400 μA, respectively, Fig. 1b reveals a very fast decay during the subsequent cycles, thus suggesting a progressive hindering of the corresponding redox process. In order to further characterize the electrochemical features of the two oxides, galvanostatic cycling tests using them as the working electrodes in lithium cell are performed at the current rate of C/10 (1C = 308 mA $g^{-1}$ for $MnO_2$ and 168 mA $g^{-1}$ for $TiO_2$). The corresponding voltage profiles reported in Fig. 1(c, d) fit with the CV curves, and show for $MnO_2$ electrode (Fig. 1c) a first cycle characterized by a single discharge plateau evolving at 2.6 V and charge processes centered at 3.3 V, respectively, that shift to 2.9 and 3.2 V in the subsequent cycles due to the above mentioned structural modification of the material and SEI film formation on the electrodes surface.[51] Relevantly, Fig. 1c evidences a very poor delivered capacity for the material with a value limited to 26 mAh $g^{-1}$ during the first discharge, decreasing to 12 mAh $g^{-1}$ during the subsequent cycles, as above suggested by the by the modest CV currents. As for the cell performance of $TiO_2$ electrode, Fig. 1d shows flat profiles typical of the biphasic (de)insertion process, with discharge taking place at 1.8 V and charge at 1.9 V,[51,52] and corresponding capacities of 120 and 80 mAh $g^{-1}$ during the first cycle. Subsequently, the delivered capacity of the Li|$TiO_2$ cell (Fig. 1d) rapidly drops to reach a value of about 27 mAh $g^{-1}$ at the 10$^{th}$ cycle, thus confirming the limits of the electrochemical process already shown by CV in Fig. 1b. Despite the low capacity values delivered by the $MnO_2$ and $TiO_2$ electrodes through cycling, the suitable stability of the electrode/electrolyte interphase in lithium cell is a very important parameter to be evaluated for a possible application of the two oxides as additives in sulfur composites. Therefore, Nyquist plots have been recorded through EIS at the open circuit voltage (OCV) condition of the lithium cells and after 10 discharge/charge cycles and reported in Fig. 1e and f. The obtained results are analyzed through the non-linear least squares (NLLS) fitting method, throughout a Boukamp software that allows the evaluation of the various resistance contributes, by building an equivalent circuit formed by resistances indicated by (R) and constant phase elements (CPE) indicated by (Q).[53,54] Table 1, including the outcomes of the NLLS analyses, reveals that the cells fit with the circuit $R_e(R_iQ_i)Q_g$ in which $R_e$ is the electrolyte resistance identified by the high-frequency intercept of the semicircle, $R_i$ in the $(R_iQ_i)$ element is the electrode/electrolyte interphase resistance corresponding to the width of the high-medium frequency semicircle with contributions of charge transfer and SEI layer, while $Q_g$ indicates the geometric capacity of the cell represented by a low frequency tilted line.[53,54] The NLLS analyses in Tab. 1 evidence relatively limited $R_i$ values for both $MnO_2$ (Fig. 1e) and $TiO_2$ (Fig. 1f), which become even lower upon cycling in lithium cell. In particular, the electrode/electrolyte interphase resistance related to $MnO_2$ drops from 186 Ω at the OCV condition to 131 Ω after 10 cycles, while that of $TiO_2$ decreases from 77 to 37 Ω. Relevantly, $TiO_2$ exhibits lower resistance values with respect to $MnO_2$ likely due to the nanometric morphology of the sample (see Figure S1 in the Supporting Information) which causally leads to a more conductive SEI with favorable ion exchange.[47] The low resistance values observed in Tab. 1 and the structural stability of the SEI suggested by the evolution of the Nyquist plots in Fig. 2(e, f) indicate a possible applicability of the two metal oxides in the sulfur composite,[55] despite the very modest contribution expected by the (de)insertion process to the overall capacity of the Li/S cell.

**Figure 1**

In summary, the analyses of the $MnO_2$ and $TiO_2$ materials displayed in Figs. S1, S2 and in Fig. 1 show different structure and morphology, as well as a stable electrode/electrolyte interphases and a poor intrinsic electrochemical contribution, thus suggesting the employment of these metal oxides in Li-S battery as an "almost inert" cathode additive, however with a possible active role in boosting the conversion reaction, increasing the lithium polysulfides retention, and mitigation their shuttle process.[32] The ability of retaining lithium polysulfides intermediates formed during Li-S conversion is qualitatively shown through photographic images in Fig. S3 (Supporting Information) and Fig. 2a, and actually evaluated by UV-vis measurements in Fig. 2b. Vials containing $MnO_2$ and $TiO_2$ samples are prepared (Fig. S3a) and filled with an electrolyte solution composed of 1,3-dioxolane (DOL) and 1,2-dimethoxyethane (DME) dissolving LiTFSI and $LiNO_3$, added with an amount of 0.5 wt.% of $Li_2S_8$ polysulfide (see preparation details in Experimental section). The pristine polysulfide-added electrolyte exhibits a dark red color due to the presence of $Li_2S_8$ (Fig. S3b, left-hand side), which is strongly mitigated upon direct contact with $MnO_2$ (Fig. S3b, center), and partly reduced by $TiO_2$ powder (Fig. S3b, right-hand side) already after 90 minutes since addition. The difference in color is further evidenced in Fig. 2a, which shows the solutions collected upon 5 hours of aging in the respective vials. The softening of the dark red color is ascribed to the partial polysulfides retention by the transition metal oxides, which is particularly effective for $MnO_2$ as suggested by the light color of the related solution.[56]. The better polysulfides retention of $MnO_2$ is confirmed by the UV-vis measurements performed on the pristine $Li_2S_8$-added electrolyte, and on solutions held in contact with $MnO_2$ and $TiO_2$ as reported in Fig. 2b. Indeed, the analyses show a characteristic signal of lithium polysulfide species in the visible region between 750 and 550 nm for all the three solutions.[57] The pristine $Li_2S_8$-added solution exhibits the expected intense response, whilst the solutions held in contact with $TiO_2$ displays a lowest polysulfide signal that almost vanishes for the one using $MnO_2$ (Fig. 2b). Therefore, the selected transition metal oxides actually reveal a lithium polysulfide retention ability, which may be influenced by the intrinsic interactions between $MnO_2$ and $TiO_2$ and $Li_2S_x$ species, as well as by the different morphology of the two samples (see Fig. S1 in Supporting Information). Consequently, a different cycling behavior of the S-$MnO_2$ and S-$TiO_2$ composites is expected in lithium cell. The structure of the S-$MnO_2$ and S-$TiO_2$ composites is determined through XRD. The resulting patterns of S-$MnO_2$ (Fig. 2c) and S-$TiO_2$ **(**Fig. 2d) exclusively show the orthorhombic sulfur ($S_8$) and the corresponding metal oxide signals, that is, $\beta$-$MnO_2$ and $\alpha$-$TiO_2$. The crystallographic data indicate the adequateness of the mild synthesis condition adopted herein for achieving a suitable physical mixing between sulfur and the metal oxide, without any side reaction that can possibly lead to the formation of additional and undesired compounds, as indeed suggested by our previous works for different composites.[58–60] The accurate determination of the sulfur content in the composites is achieved by the thermogravimetric analyses (TGA) and the corresponding differentiate thermogravimetry (DTG) reported in Fig. 2e for S-$MnO_2$ and Fig. 2f for S-$TiO_2$. Indeed, the measurements exhibit a single weight drop centered at 300 °C ascribable to sulfur evaporation[61] that corresponds to a sulfur loading of 80 % in both S-$MnO_2$ and S-$TiO_2$. Thus, the analyses evidence that the facile synthesis pathway adopted herein avoid sulfur loss through evaporation and allows the tuning of high amounts of sulfur in composites designed for high energy density lithium batteries.

**Figure 2**

This synthetic pathway, that may be reasonably scaled to achieve practical production, does not affect the metal oxides morphology, and lead to their uniform distribution into the composites as determined by SEM coupled with X-ray energy dispersive spectroscopy (EDS) (Fig. 3a-j). Indeed, the SEM-EDS analyses reveal large sulfur domains (up to 20 µm) homogenously decorated with micrometric $MnO_2$ (Fig. 3a-e), nanosized $TiO_2$ (Fig. 3f-j), that is, sulfur including metal oxides with similar morphology to that of bare materials observed in Fig. S1. The main morphological difference observed for the two composites, which is principally ascribed to the additives, is reasonably expected to differently influence their electrochemical behavior in Li-S battery as will be shown hereafter.

**Figure 3**

The composites are then employed in lithium cell and investigated through CV and EIS as reported in Figure 4. The voltammograms related to the Li|S-$MnO_2$ and Li|S-$TiO_2$ cells are displayed in Fig. 4a and Fig. 4b, while the corresponding Nyquist plots acquired through EIS at the OCV condition and after 1, 5 and 10 CV cycles are displayed in Fig. 4c and Fig. 4d, respectively. The first CV cycle shows for both the S-metal oxide electrodes a reversible conversion process, indicated by the two reduction peaks at about 2.28 and 2.0 V *vs.* $Li^+$/Li during the cathodic scan, which are reversed into a double oxidation signal extending from 2.3 to 2.5 V *vs.* $Li^+$/Li during the subsequent anodic scan. This response is in line with the multi-step electrochemical process leading to the formation of lithium polysulfides with various chain length during discharge, namely, $Li_2S_8$, $Li_2S_6$ and subsequently $Li_2S_x$ with $2 \leq x \leq 4$, which are conversed back to Li and S upon charge.[62] It is worth noting that the narrow peaks exhibited by the two voltammograms, and the overlapping of the subsequent potential profiles actually suggest adequate kinetics and stability of the electrochemical reaction, despite the slight decrease of the peak current upon scanning may imply minor losses of active material. Furthermore, the exclusive presence of the potential signature ascribed to the Li-S conversion process, without any additional CV peak, confirms the negligible contribute of the electrochemical (de)insertion reaction of $MnO_2$ and $TiO_2$, as already speculated during the discussion of Fig. 4. On the other hand, the shift of the discharge signal centered at 2.28 V *vs.* $Li^+$/Li to higher potential values, that is, 2.33 V *vs.* $Li^+$/Li for S-$MnO_2$ (Fig. 4a) and 2.31 V *vs.* $Li^+$/Li for S-$TiO_2$ (Fig. 4b) observed upon the first cycle indicates the decrease of the discharge/charge polarization. This *activation* is associated with the improvement of the electrode/electrolyte interphase conductivity due to the stabilization of an adequate SEI layer, and the advantageous migration by electrodeposition of the insulating sulfur from the surface of the porous carbon collector to its bulk which, in turn, enhances the electrical contact of the active material with the conductive matrix and promotes the ion transport in the interphase as demonstrated in previous works.[59,60] The increase of electrode conductivity is well confirmed by the Nyquist plots recorded upon CV (Fig. 4c, d), which are analyzed by NLLS fitting method and the results reported in Table 2.[53,54] Both Li-S cells show a decrease of the interphase resistance ($R_i$) upon cycling, from values of 24 Ω for S-$MnO_2$ and 14 Ω for S-$TiO_2$ at the OCV condition, to 6 Ω and 5 Ω, respectively, after 10 CV cycles. On the other hand, the NLLS analyses of Tab. 2 reveal differences in the equivalent circuits of the two Li-S systems. Indeed, the Nyquist plots the Li|S-$TiO_2$ cell (Fig. 4d) show at low frequencies a line tilted at about 45° which is typical of a Warburg-type semi-infinite $Li^+$ diffusion, indicated by the CPE ($Q_w$) in the corresponding equivalent circuit.[63] Instead, the low frequency region of the Li|S-$MnO_2$ cell (Fig. 4c) evidences the presence of an almost vertical line attributed to the geometric capacity in a quasi-blocking electrode

setup, indicated by the CPE ($Q_g$) in the corresponding equivalent circuit of Tab. 2. It is worth mentioning that the number of the ($R_iQ_i$) elements in the circuits of Tab. 1 varies by the ongoing of the CV test due to the progressive and favorable modification of the electrode/electrolyte interphase.[59,60]

**Figure 4**

Further electrochemical characteristics of the two materials are provided by the rate capability tests in Figure 5, carried out on the Li|S-$MnO_2$ and Li|S-$TiO_2$ cells at the increasing currents of C/10, C/8, C/5, C/3, C/2, 1C and 2C (1C = 1675 mA $g_S^{-1}$). The voltage profile of S-$MnO_2$ (Fig. 5a) and of S-$TiO_2$ (Fig. 5b) show at C/10 two distinct discharge plateaus evolving at about 2.3 and 2.1 V and charge plateaus merging between 2.2 and 2.4 V, thus accounting for the reversible conversion of Li and S to lithium polysulfides ($Li_2S_x$ with $2 \leq x \leq 8$) in full agreement with the CV response discussed in Fig. 4. Moreover, the voltage profiles reveal a low polarization between the discharge and charge processes that slightly grows at higher C-rates due to the increasing ohmic drop triggered by raising the current. In spite of the similarity and the adequateness of the Li|S-$MnO_2$ and the Li|S-$TiO_2$ cell responses at current ranging from C/10 to 1C, the S-$MnO_2$ electrode reveals a poor behavior at the relatively high current rate of 2C (Fig. 5a), instead the S-$TiO_2$ well operates at the same C-rate (Fig. 5b). This aspect is further evidenced by the corresponding cycling trends: indeed, S-$MnO_2$ (Fig. 5c) delivers capacity values of 1060, 1030, 990, 930, 870 and 780 mAh $g_S^{-1}$ at the C/10, C/8, C/5, C/3, C/2 and 1C rates that drops to 160 mAh $g_S^{-1}$ at 2C, while S-$TiO_2$ shows a discharge capacity of 1100, 1060, 1030, 990, 940 and 870 mAh $g_S^{-1}$ from C/10 to 1C, and still delivers a value of 790 mAh $g_S^{-1}$ at 2C, thus suggesting a better rate capability. The above difference may be attributed to a faster kinetics of the Li-S conversion promoted by the nanometric $TiO_2$ included in the latter composite with respect to the micrometric $MnO_2$ in the former one. It is worth mentioning that the notable lithium polysulfides retention of the $MnO_2$ may possibly limit the kinetics of the Li-S conversion and rate capability of the cell,[24] that is in line with the better rate capability of the S-$TiO_2$ electrode observed in Fig. 5. In addition, the nanometric morphology of the $TiO_2$ additive (Fig. S1 in Supporting Information) may further boost the electrode kinetics and the cycling rate of the cell. On the other hand, both materials recover almost completely the pristine electrochemical response upon lowering the C-rate back to C/10 after 35 cycles, as S-$MnO_2$ delivers 995 mAh $g_S^{-1}$ and S-$TiO_2$ 1050 mAh $g_S^{-1}$ that correspond to about 95 % of the respective initial capacity. These notable performances indicate an outstanding capacity retention upon the stress caused by raising the current, likely ascribed to the polysulfide-retaining ability of the metal oxides in the composite sulfur cathode and particularly favored by the nanometric morphology.[32,35,64]

**Figure 5**

Figure 6 reports the cycling trends related to the prolonged galvanostatic cycling tests at fixed current rates, that is, C/5 (Fig. 6a), 1C (Fig. 6b) and 2C (Fig. 6c), performed to evaluate the cycle life of the Li-S cells. The two sulfur-metal oxide composites show adequate cycle life, extended over 120 cycles at the lower rates C/5 (Fig. 6a) and 1C (Fig. 6b), with promising delivered capacity values, suitable capacity retention, and high coulombic efficiency. Furthermore, the corresponding voltage profiles displayed in Figure S4 in Supporting Information reveal the reversible Li-S conversion process and a low polarization of the discharge and charge processes at C/5 (Fig. S4a, b) and 1C (Fig. S4c, d) for S-$MnO_2$ and S-$TiO_2$, respectively. In particular, the sulfur composites exhibit similar behavior at C/5 (Fig. 6a) with an initial capacity around 1120 mAh $g_S^{-1}$ retained for the 69 % using S-$MnO_2$ and for 74 % using S-$TiO_2$ upon the 120 cycles taken into account, and coulombic efficiencies of 99 % after the initial stages of the tests. Instead, at 1C (Fig. 6b) S-$MnO_2$ displays a slightly higher initial capacity with respect to S-$TiO_2$, that is, 1140 and 930 mAh $g_S^{-1}$ after the first cycle. However, S-$MnO_2$ evidences a faster capacity decay during the cycling test and shows a final capacity of 700 mAh $g_S^{-1}$ that corresponds to 61 % of the initial value, while S-$TiO_2$ retains its capacity for 76 %. Moreover, S-$TiO_2$ shows a relevant cell life of 400 cycles at the relatively high current rate of 2C (Fig. 6c an d Fig. S4f), with an initial capacity of 850 mAh $g_S^{-1}$ retained for the 52

% at the end of the test, and a final coulombic efficiency of 98 %. Instead, S-$MnO_2$ shows at 2C a very poor electrochemical response (Fig. S4e) in agreement with the rate capability outcomes discussed in Fig. 5, thus further suggesting the faster reaction evolution promoted by the nanometric $TiO_2$ compared to the micrometric $MnO_2$, which may also favor an excessive lithium polysulfides retention as discussed in Fig. 5. The small discrepancies of the delivered capacity values observed for rate capability tests (Fig. 5) and the long-term cycling (Fig. 6) may be due to slight difference in sulfur loading exploited by the S-$MnO_2$ and S-$TiO_2$ electrodes (i.e., 2.0 mg $cm^{-2}$ for rate capability tests and between 1.5 and 2.0 mg $cm^{-2}$ for long-term cycling tests) and to the adopted operative conditions. Indeed, the rate capability evaluation involves initial cycling at low C-rates that may influence kinetics and activation of the Li-S process (see experimental section). In summary, the galvanostatic tests in lithium cell suggest the direct physical combination of sulfur and metal oxides as viable synthetic strategy to achieve satisfactory cyclability of Li-S cells, and evidence the better response of S-$TiO_2$ with respect to S-$MnO_2$ due to the different morphology of the metal oxide additive.

**Figure 6**

The achievement of Li-S battery of practical interest certainly requires a high sulfur loading and a low electrolyte/sulfur (E/S) ratio to reach adequate energy density for a possible commercialization and enhance the cell characteristics compared to the state-of-the-art battery.[65–67] In this regard, Figure 7 displays additional galvanostatic tests performed on the Li-S cells employing electrodes with sulfur loading raised from about 2 mg $cm^{-2}$ to about 5 to 6 mg $cm^{-2}$ (referred to electrode geometric area) and E/S ratio in the cell lowered from 15 µl $mg^{-1}$ to 10 µl $mg^{-1}$ with respect to the tests reported above (Fig. 5 and Fig. 6). The measurements are carried out at a C-rate of C/5, which corresponds to a current density approaching 2 mA $cm^{-2}$, with a sulfur loading of 5.4 mg $cm^{-2}$ for S-$MnO_2$ and 5.8 mg $cm^{-2}$ for S-$TiO_2$. The resulting voltage profiles reveal reversible and lowly polarized discharge for both S-$MnO_2$ (Fig. 7a) and S-$TiO_2$ (Fig. 7b), and a charge processes exhibiting well-defined plateaus for the conversion of lithium polysulfides to Li and S. Interestingly, the discharge profiles in Fig. 7a and b show a sloped trend that may be ascribed to the hindering of the $Li^+$ diffusion at the electrode/electrolyte interphase caused by the relevant sulfur content and by the relatively high current density.[60,68] Nevertheless, the Li-S cells deliver over 90 cycles with a coulombic efficiency approaching 100 % after the initial stage (insets in Fig. 7a and b) and steady state capacity exceeding 900 mAh $g_S^{-1}$ that corresponds to about 5 mAh $cm^{-2}$, with a final value higher than 4.5 mAh $cm^{-2}$, as revealed by the cycling trends reported in Fig. 7c for S-$MnO_2$ and Fig. 7d for S-$TiO_2$. In particular, the S-$MnO_2$ electrode with a sulfur loading of 5.4 mg $cm^{-2}$ exhibits an initial areal capacity of 5.9 mAh $cm^{-2}$ (1090 mAh $g_S^{-1}$) which decreases to 5.3 mAh $cm^{-2}$ (990 mAh $g_S^{-1}$) after 10 cycles and to 4.7 mAh $cm^{-2}$ (865 mAh $g_S^{-1}$) after 90 cycles, while the S-$TiO_2$ electrode with a sulfur loading of 5.8 mg $cm^{-2}$ delivers 5.7 mAh $cm^{-2}$ (985 mAh $g_S^{-1}$), slightly decreasing to 5.6 mAh $cm^{-2}$ (970 mAh $g_S^{-1}$) upon the first 10 cycles and to 4.7 mAh $cm^{-2}$ (810 mAh $g_S^{-1}$) at the end of the test. In summary, both the sulfur composites demonstrate promising performance even in a challenging conditions such as that providing an increased sulfur loading and a limited E/S ratio, with a notable capacity retention attributed to the restraining action of the metal oxides into the cathode, and the limited polysulfides shuttle in the optimized cells.[69] In this regard, control electrodes exploiting exclusively sulfur were tested in lithium cells to evaluate the possible differences in the electrochemical behavior due to the absence of the metal oxides. Figure S5 (Supporting Information) displays the results of cycling tests carried out on Li cells employing the above sulfur control electrodes with loading of either 5.0 (Fig. S5a, c) or 4.5 mg $cm^{-2}$ (Fig. S5b, d). Both the galvanostatic profiles (Fig. S5a, b) and the related cycling trends (Fig. S5c, d) reveal satisfactory initial performance comparable to those exhibited by the S-$MnO_2$ and S-$TiO_2$ electrodes (compare Fig. S5 with Fig. 7) which can be attributed to the electrode formulation adopted herein exploiting the carbon-cloth GDL support (see Experimental section). Indeed, this substrate is characterized by relevant porosity that ensures an optimal electric contact of the active material with the electrode structure and allows fast kinetics of the Li-S conversion process.[60] However, both cells exhibit short cycle life and scarce rate capability due to several short circuits ascribed to massive formation of dendrites on the Li surface.

This behavior is likely due to the massive diffusion of the lithium polysulfides during Li-S operation that reach and directly react with the anode favoring the formation of an irregular SEI and the uneven deposition of $Li^+$ ions during charge.[70] This phenomenon can be strongly mitigated by the addition of oxides that can retain the polysulfides and hinder their fast migration to the anode such as $MnO_2$ and $TiO_2$, thus allowing safe cycling even at a sulfur loading exceeding 5 mg $cm^{-2}$ as demonstrated in Fig. 7. These data may actually shed light on the various sustainable strategies for scaling-up efficient configurations of high-energy lithium-sulfur cells with practical interest.[42,59]

**Figure 7**

## Conclusions

Sulfur composites synthetized by a simple pathway, composed of sulfur (80 wt.%) and either $MnO_2$ or $TiO_2$ powder (20 wt.%) were investigated for application in lithium battery. XRD measurements revealed a well-defined $\beta$ and $\alpha$ phases for $MnO_2$ and $TiO_2$ powders, respectively, while SEM images displayed micrometric aggregates for $MnO_2$ and nanometric particles for $TiO_2$. Cyclic voltammetry (CV) and galvanostatic cycling tests performed in lithium cells shown modest side electrochemical activity for the $MnO_2$ electrode and fast deactivation of the $TiO_2$, and suggested a negligible contribution of the (de)insertion processes into the metal oxides to the overall capacity of the lithium-sulfur conversion reaction. Moreover, electrochemical impedance spectroscopy (EIS) carried out upon galvanostatic tests of the oxides in lithium cell exhibited a decrease of the electrode/electrolyte interphase resistance by cycling, thus accounting for formation of a stable solid electrolyte interphase (SEI) and confirming the applicability of the $MnO_2$ and $TiO_2$. XRD and SEM-EDS analyses of the S-$MnO_2$ and S-$TiO_2$ composites exhibited the retention of the oxides structure and morphology, the absence of significant impurities and the homogeneous distribution of the metal oxide in the sulfur matrices. Furthermore, CV measurements of Li|S-$MnO_2$ and Li|S-$TiO_2$ cells evidenced a reversible, stable and lowly polarized Li-S conversion process characterized by two

separated discharge signals at 2.28 and 2.0 V *vs.* $Li^+$/Li, and a broad charge double-signal between 2.3 and 2.5 V *vs.* $Li^+$/Li, thus excluding side electrochemical activity of the $MnO_2$ and $TiO_2$. The voltammograms also revealed an *activation* of the Li-S process suggested by the shift of the discharge signal from 2.28 V *vs.* $Li^+$/Li to values exceeding 2.3 V *vs.* $Li^+$/Li, and the remarkable decrease of the electrode/electrolyte interphase resistance confirmed by EIS carried out upon CV. In particular, the resistance values of 24 Ω and 14 Ω measured at the OCV condition for the Li|S-$MnO_2$ and Li|S-$TiO_2$ systems, respectively, decreased to about 5 and 6 Ω after 10 CV cycles. Moreover, the EIS data revealed a Warburg-type $Li^+$ diffusion for the S-$TiO_2$ at the charged state, and a blocking-electrode trend for S-$MnO_2$, as likely ascribed to the different morphology of the two oxides in the composites which also influenced their cycling behavior. Indeed, rate capability tests displayed an efficient Li-S conversion process according with the CV data for the two sulfur composites from C/10 to 1C current, whilst only S-$TiO_2$ demonstrated a suitable cycling at 2C. Moreover, S-$TiO_2$ exhibited better capacity retention, despite both the Li-S cells recovered over the 90 % of the initial capacity when the current was lowered back to C/10. Prolonged galvanostatic tests confirmed the better response of the material using the nanometric $TiO_2$ and confirmed the kinetic limit related to the micrometric $MnO_2$. Hence, the Li|S-$MnO_2$ and Li|S-$TiO_2$ systems shown satisfactory trends over 120 cycles at C/5 and 1C rate, with a higher capacity retention for the S-$TiO_2$ electrode, particularly at 1C. Furthermore, only S-$TiO_2$ delivered an adequate capacity at 2C current, with a value of 850 mAh $g_S^{-1}$ retained for over 50 % after 400 discharge/charge cycles. In order to achieve more scalable Li-S cells, the sulfur loading in the electrode was increased up to 5.4 mg $cm^{-2}$ for S-$MnO_2$ and 5.8 mg $cm^{-2}$ for S-$TiO_2$, and the electrolyte/sulfur (E/S) ratio was limited to 10 µl $mg^{-1}$. The related galvanostatic tests at the current rate of C/5 (about 2 mA $cm^{-2}$) revealed initial areal capacity approaching 6 mAh $cm^{-2}$ (as normalized to the electrode geometric surface) and a notable final value of about 4.7 mAh $cm^{-2}$ after 90 discharge/charge cycles, during which S-$TiO_2$ material exhibited once more a better retention. Therefore, the experimental outcomes of this work suggested the physical mixing pathway of sulfur and metal oxides as a practical process to achieve a composite for efficient Li-S battery, and evidenced the benefits of a nanometric morphology of the additive to allow fast reaction kinetics, long cycle life and satisfactory capacity retention.

## Experimental

### *Metal oxides*

The structure and morphology of manganese (IV) oxide ($MnO_2$, ≥ 99 %, Sigma-Aldrich) and titanium (IV) oxide ($TiO_2$, anatase, 99.8 % trace metal basis, Sigma-Aldrich) powders were analyzed by X-ray diffraction (XRD) and scanning electron microscopy (SEM). The XRD patterns were acquired via a Bruker D8 Advance diffractometer equipped with a Cu-Kα source (8.05 keV) by performing scans between 10 and 90° (2θ) at 10 s step$^{-1}$ with a step size of 0.02°. A Zeiss EVO MA10 exploiting a tungsten thermionic electron source was used to collect SEM images, while TEM analyses were carried out though a Zeiss EM 910 microscope equipped with a tungsten thermoionic electron gun operating at 100 kV. Previous to TEM observation, the $MnO_2$ and $TiO_2$ samples were dispersed in water and sonicated through a Branson 3200 ultrasonic cleaner.

The electrochemical features of the $MnO_2$ and $TiO_2$ powders were preliminarily evaluated in lithium cell. $MnO_2$ and $TiO_2$ electrodes were prepared through solvent casting technique by using a doctor blade (MTI Corp.). The electrode slurries were prepared by mixing either $MnO_2$ or $TiO_2$ powders with Super P carbon (SPC, Timcal) conductive agent and polyvinylidene fluoride (Solef ® 6020 PVDF) polymeric binder in the 80:10:10 weight ratio by using the *N*-methyl-2-pyrrolidone (NMP, Sigma-Aldrich) solvent. The slurries were cast on carbon-coated aluminum foils (MTI Corp.), heated at 70 °C for 3 hours to remove the NMP and subsequently cut into 10 mm-electrode disks which were dried overnight under vacuum at 110 °C to ensure the removal of water or solvent traces. The metal oxide electrodes were then transferred in an Ar-filled MBraun glovebox ($H_2O$ and $O_2$ content below 1 ppm) to assemble lithium T-type cells. Three-electrode cells were obtained by combining a working metal oxide electrode (either $MnO_2$ or $TiO_2$) with two 10 mm-diameter lithium disks acting as counter and reference electrodes, respectively. Working and counter electrodes were separated by a glass fiber 10 mm-diameter disk (GF/B Whatman ®) soaked with the electrolyte, that is, a solution of ethylene carbonate (EC) and dimethyl carbonate (DMC) mixed in the 1:1 volume ratio dissolving the lithium hexafluorophosphate ($LiPF_6$) conducting salt in concentration of 1 mol $dm_{solvent}^{-3}$. The electrolyte solution was purchased by Sigma-Aldrich (LP-30) and is indicated as EC:DMC, 1M $LiPF_6$. Two-electrode cells were prepared with the same method by excluding the lithium reference disk.

Cyclic voltammetry (CV) measurements were performed on three-electrode T-cells at the scan rate of 0.1 mV $s^{-1}$ in the 1.9 – 3.8 V *vs*. $Li^+$/Li and 1.4 – 2.8 V *vs*. $Li^+$/Li potential ranges for $MnO_2$ and $TiO_2$, respectively. Galvanostatic cycling tests were carried out at the constant current rate of C/10 (1C = 308 mA $g^{-1}$ for $MnO_2$ and 1C = 168 mA $g^{-1}$ for $TiO_2$) in two-electrode lithium T-cells. A voltage limit of 1.9 – 3.8 V was used for the $MnO_2$ cell, while the $TiO_2$ one was cycled between 1.4 and 2.8 V. Electrochemical impedance spectroscopy (EIS) was performed on the two-electrode cells at the open circuit voltage (OCV) condition and upon 10 discharge/charge cycles at C/10 in the 500 kHz – 100 mHz frequency range by using a 10 mV alternate voltage signal.

*Electrolyte preparation*

The electrolyte was prepared by mixing 1,3-dioxolane (DOL, anhydrous, contains ca. 75 ppm BHT as inhibitor, 99.8 %, Sigma-Aldrich) and 1,2-dimethoxyethane (DME, anhydrous, 99.5 %, inhibitor-free, Sigma-Aldrich) with a 1:1 weight ratio, and dissolving lithium bis(trifluoromethanesulfonyl)imide ($LiN(SO_2)_2(CF_3)_2$, LiTFSI, 99.95 % trace metals basis, Sigma-Aldrich) as conductive salt and lithium nitrate ($LiNO_3$, 99.99 % trace metals basis, Sigma-Aldrich) as passivating agent in the solvents mixture in a 1 mol $kg_{solvent}^{-1}$ concentration for each salt. Before employment, DOL and DME solvents were stored under molecular sieves (rods, 3 Å, size 1/16 in., Honeywell Fluka) until a water content lower than 10 ppm was achieved as determined by a Karl Fischer 899 Coulometer (Metrohm), while LiTFSI and $LiNO_3$ were dried for 2 days under vacuum at 110 °C. The electrolyte solution is indicated as DOL:DME, 1 mol $kg^{-1}$ LiTFSI, 1 mol $kg^{-1}$ $LiNO_3$.

An additional electrolyte containing polysulfide for UV-vis analyses was prepared by adding 0.5 wt.% of $Li_2S_8$ to the DOL:DME, 1 mol kg$^{-1}$ LiTFSI, 1 mol kg$^{-1}$ $LiNO_3$ solution. The $Li_2S_8$ addition procedure is reported in a previous work.[71]

UV-vis analyses were carried out on the above DOL:DME, 1 mol kg$^{-1}$ LiTFSI, 1 mol kg$^{-1}$ $LiNO_3$, 0.5 wt.% $Li_2S_8$ samples upon 5 hours aging in contact with $MnO_2$ and $TiO_2$ powders in the 500 – 800 nm region. Absorption spectra were collected with an Agilent Cary 300 UV-Vis spectrophotometer at room temperature against a DOL:DME 1:1 *w/w* reference solution.

*Sulfur composites*

Sulfur (≥ 95 %, Riedel-de Haën) and either $MnO_2$ (≥ 99 % Sigma-Aldrich) or $TiO_2$ (99.8 % Sigma-Aldrich) powders were mixed in the 80:20 weight ratio by magnetic stirring at 125 °C until complete melting of sulfur and homogenization. The value of 125 °C was selected to ensure the complete melting of sulfur that starts at about 113 °C, and, at the same time, avoid possible evaporation with lowering of the sulfur content in the composite that may occur at higher temperatures. The sulfur-metal composites were subsequently achieved by cooling down the mixture to room temperature, and grinding the resulting solid to get a fine powder. The composites, that is, S:$MnO_2$ 80:20 w/w and S:$TiO_2$ 80:20 w/w, are subsequently indicated by the acronyms S-$MnO_2$ and S-$TiO_2$, respectively.

Structure and morphology of the sulfur composites were investigated in the same conditions as for the bare metal oxides, that is, by XRD using Bruker D8 Advance diffractometer equipped with a Cu-Kα source (8.05 keV) scanning in the 10 – 90° 2θ range with a 10 s step$^{-1}$ scan rate and 0.02° step, and by SEM employing a Zeiss EVO MA10 using a tungsten thermionic electron source. Elements distribution analysis was carried out on the SEM images through energy dispersive X-ray spectroscopy (EDS) via a X-ACT Cambridge Instruments analyzer.

Thermogravimetric analyses (TGA) were carried out on S-$MnO_2$ and S-$TiO_2$ samples in the 25 – 800 °C temperature range under $N_2$ flow at a rate of 5 °C min$^{-1}$ through a Mettler-Toledo TGA 2 instrument.

S-$MnO_2$ and S-$TiO_2$ electrodes were prepared through doctor blade (MTI Corp.) casting of slurries composed by the sulfur composite, SPC (Timcal), and PVDF (Solef ®) with the 80:10:10 weight ratio dispersed in NMP (Sigma-Aldrich). The slurries composition allows a final sulfur content as high as 64 wt.% in the cathodes. Sulfur control electrodes were also prepared through a slurry composed by elemental sulfur, SPC and PVDF in the 80:10:10 weight ratio. The electrode slurries were coated on a carbon-cloth gas diffusion layer (GDL ELAT 1400, MTI Corp.), heated at 50 °C to remove the NMP, cut into 14-mm disks and dried under vacuum at 40 °C overnight. Subsequently, the electrodes were transferred in an Ar-filled MBraun glovebox ($H_2O$ and $O_2$ content below 1 ppm) to assemble CR2032 coin-type cells (MTI Corp.) with 14 mm-diameter lithium disk anode separated from the sulfur cathode through a 16 mm-diameter Celgard 2400 foil soaked with the DOL:DME, 1 mol $kg^{-1}$ LiTFSI, 1 mol $kg^{-1}$ $LiNO_3$ electrolyte solution. The electrolyte/sulfur (E/S) ratios used for the electrochemical tests are specified below.

*Electrochemical measurements*

CV, EIS and galvanostatic cycling measurements were carried out in lithium coin-cells exploiting sulfur cathodes, initially with an active material loading between 1.5 and 2 mg $cm^{-2}$ (referred to the electrode geometric area of 1.54 $cm^2$) and an electrolyte to sulfur (E/S) ratio of 15 µL $mg^{-1}$. CV scans were performed in the 1.8 – 2.8 V *vs.* $Li^+$/Li potential range at 0.1 mV $s^{-1}$, while EIS was carried out at the OCV cell condition and upon voltammetry after 1, 5 and 10 cycles within the 500 kHz – 100 mHz frequency range by applying a 10 mV alternate voltage signal. The rate capability of the cells was evaluated through galvanostatic tests at a C-rate increasing every 5 cycles, that is, C/10, C/8, C/5, C/3, C/2, 1C and 2C and decreasing back at C/10 after the 35$^{th}$ cycle, while the cells cycle life was investigated by prolonged cycling tests at C/5, 1C and 2C constant current rates (1C = 1675 mA $g_S^{-1}$). Discharge and charge processes were limited between 1.9 and 2.8 V from C/10 to C/2 and between 1.8 and 2.8 V at 1C and 2C. Additional galvanostatic cycling measurements were carried out at C/5 between 1.7 and 2.8 V in lithium cells employing sulfur cathodes with active material loading (referred to the electrode geometric area of 1.54 $cm^2$) of 5.4 and 5.8 mg $cm^{-2}$ for S-$MnO_2$ and S-$TiO_2$, respectively and an E/S ratio of 10 µL $mg^{-1}$.

The sulfur control electrodes were tested in lithium coin-cells through galvanostatic cycling measurements at the constant rate of C/5 in the 1.7 – 2.8 V voltage range. The cells employed control electrodes with sulfur loading of 5.0 and 4.5 mg $cm^{-2}$ and E/S ratio of 10 µL $mg^{-1}$.

All the CV and EIS measurements were carried out via a VersaSTAT MC Princeton Applied Research (PAR–AMETEK) analyzer, while a MACCOR series 4000 battery test system was employed for the galvanostatic cycling tests. The Nyquist plots recorded by EIS were analyzed with a Boukamp tool by applying the non-linear least squares (NLLS) fitting method to extrapolate equivalent circuits and resistance values.[53,54] Only fits exhibiting a $\chi^2$ value of about $10^{-4}$ or lower were considered acceptable.

**Acknowledgements**

This project/work has received funding from the European Union's Horizon 2020 research and innovation programme Graphene Flagship under grant agreement No 881603. The authors also thank grant "Fondo di Ateneo per la Ricerca Locale (FAR) 2021", University of Ferrara, and the collaboration project "Accordo di Collaborazione Quadro 2015" between University of Ferrara (Department of Chemical and Pharmaceutical Sciences) and Sapienza University of Rome (Department of Chemistry).

## References

[1] N. O. Bonsu, *Journal of Cleaner Production* **2020**, *256*, 120659.

[2] X. Wang, Y. Ding, Y. Deng, Z. Chen, *Advanced Energy Materials* **2020**, *10*, 1903864.

[3] S. Chen, F. Dai, M. Cai, *ACS Energy Letters* **2020**, *5*, 3140.

[4] D. Di Lecce, R. Verrelli, J. Hassoun, *Green Chemistry* **2017**, *19*, 3442.

[5] M. A. Rajaeifar, P. Ghadimi, M. Raugei, Y. Wu, O. Heidrich, *Resources, Conservation and Recycling* **2022**, *180*, 106144.

[6] L. Carbone, S. G. Greenbaum, J. Hassoun, *Sustainable Energy & Fuels* **2017**, *1*, 228.

[7] B. Scrosati, J. Hassoun, Y.-K. Sun, *Energy & Environmental Science* **2011**, *4*, 3287.

[8] D. di Lecce, V. Marangon, H.-G. Jung, Y. Tominaga, S. Greenbaum, J. Hassoun, *Green Chemistry* **2022**, DOI 10.1039/D1GC03996B.

[9] X. Ji, K. T. Lee, L. F. Nazar, *Nature Materials* **2009**, *8*, 500.

[10] A. Manthiram, Y. Fu, S.-H. Chung, C. Zu, Y.-S. Su, *Chemical Reviews* **2014**, *114*, 11751.

[11] H. Wang, Y. Yang, Y. Liang, J. T. Robinson, Y. Li, A. Jackson, Y. Cui, H. Dai, *Nano Letters* **2011**, *11*, 2644.

[12] W. Ren, W. Ma, S. Zhang, B. Tang, *Energy Storage Materials* **2019**, *23*, 707.

[13] L. E. Camacho-Forero, T. W. Smith, S. Bertolini, P. B. Balbuena, *Journal of Physical Chemistry C* **2015**, *119*, 26828.

[14] G. Zhou, E. Paek, G. S. Hwang, A. Manthiram, *Advanced Energy Materials* **2016**, *6*, 1501355.

[15] S. S. Zhang, *Inorganic Chemistry Frontiers* **2015**, *2*, 1059.

[16] M. Inagaki, M. Toyoda, Y. Soneda, T. Morishita, *Carbon N Y* **2018**, *132*, 104.

[17] L. Xiao, Y. Cao, J. Xiao, B. Schwenzer, M. H. Engelhard, L. V. Saraf, Z. Nie, G. J. Exarhos, J. Liu, *Advanced Materials* **2012**, *24*, 1176.

[18] G. Zhou, H. Tian, Y. Jin, X. Tao, B. Liu, R. Zhang, Z. W. Seh, D. Zhuo, Y. Liu, J. Sun, J. Zhao, C. Zu, D. S. Wu, Q. Zhang, Y. Cui, *Proceedings of the National Academy of Sciences* **2017**, *114*, 840.

[19] Z. Sun, J. Zhang, L. Yin, G. Hu, R. Fang, H.-M. Cheng, F. Li, *Nature Communications* **2017**, *8*, 14627.

[20] Y. Zheng, S. Zheng, H. Xue, H. Pang, *Journal of Materials Chemistry A* **2019**, *7*, 3469.

[21] W. Chen, T. Lei, T. Qian, W. Lv, W. He, C. Wu, X. Liu, J. Liu, B. Chen, C. Yan, J. Xiong, *Advanced Energy Materials* **2018**, *8*, 1702889.

[22] A. Vizintin, R. Guterman, J. Schmidt, M. Antonietti, R. Dominko, *Chemistry of Materials* **2018**, *30*, 5444.

[23] I. Gracia, H. Ben Youcef, X. Judez, U. Oteo, H. Zhang, C. Li, L. M. Rodriguez-Martinez, M. Armand, *Journal of Power Sources* **2018**, *390*, 148.

[24] V. Marangon, D. Di Lecce, L. Minnetti, J. Hassoun, *ChemElectroChem* **2021**, *8*, 3971.

[25] S. S. Zhang, *Electrochimica Acta* **2012**, *70*, 344.

[26] H. Zhang, G. G. Eshetu, X. Judez, C. Li, L. M. Rodriguez-Martínez, M. Armand, *Angewandte Chemie International Edition* **2018**, *57*, 15002.

[27] A. Rosenman, R. Elazari, G. Salitra, E. Markevich, D. Aurbach, A. Garsuch, *Journal of The Electrochemical Society* **2015**, *162*, A470.

[28] Y.-S. Su, A. Manthiram, *Nature Communications* **2012**, *3*, DOI 10.1038/ncomms2163.

[29] G. Ma, Z. Wen, Q. Wang, C. Shen, P. Peng, J. Jin, X. Wu, *Journal of Power Sources* **2015**, *273*, 511.

[30] J. He, Y. Chen, A. Manthiram, *Energy & Environmental Science* **2018**, *11*, 2560.

[31] W. Hu, Y. Hirota, Y. Zhu, N. Yoshida, M. Miyamoto, T. Zheng, N. Nishiyama, *ChemSusChem* **2017**, *10*, 3557.

[32] X. Liu, J.-Q. Huang, Q. Zhang, L. Mai, *Advanced Materials* **2017**, *29*, 1601759.

[33] X. Liang, C. Hart, Q. Pang, A. Garsuch, T. Weiss, L. F. Nazar, *Nature Communications* **2015**, *6*, DOI 10.1038/ncomms6682.

[34] M. Li, Y. Dai, X. Pei, W. Chen, *Applied Surface Science* **2022**, *579*, 152178.

[35] X. Liang, L. F. Nazar, *ACS Nano* **2016**, *10*, 4192.

[36] A. Benítez, J. Amaro-Gahete, Y.-C. Chien, Á. Caballero, J. Morales, D. Brandell, *Renewable and Sustainable Energy Reviews* **2022**, *154*, 111783.

[37] V. Marangon, C. Hernández-Rentero, M. Olivares-Marín, V. Gómez-Serrano, Á. Caballero, J. Morales, J. Hassoun, *ChemSusChem* **2021**, *14*, 3333.

[38] A. Y. S. Eng, V. Kumar, Y. Zhang, J. Luo, W. Wang, Y. Sun, W. Li, Z. W. Seh, *Advanced Energy Materials* **2021**, *11*, 2003493.

[39] Y. Ye, F. Wu, S. Xu, W. Qu, L. Li, R. Chen, *The Journal of Physical Chemistry Letters* **2018**, *9*, 1398.

[40] Y. Ge, P. Chen, W. Zhang, Q. Shan, Y. Fang, N. Chen, Z. Yuan, Y. Zhang, X. Feng, *New Journal of Chemistry* **2020**, *44*, 11365.

[41] J. Guo, S. Zhao, Y. Shen, G. Shao, F. Zhang, *ACS Sustainable Chemistry & Engineering* **2020**, *8*, 7609.

[42] A. Benítez, Á. Caballero, E. Rodríguez-Castellón, J. Morales, J. Hassoun, *ChemistrySelect* **2018**, *3*, 10371.

[43] D. Di Lecce, V. Marangon, W. Du, D. J. L. Brett, P. R. Shearing, J. Hassoun, *Journal of Power Sources* **2020**, *472*, 228424.

[44] A. Bolzan, C. Fong, B. Kennedy, C. Howard, *Australian Journal of Chemistry* **1993**, *46*, 939.

[45] I. Djerdj, A. M. Tonejc, *Journal of Alloys and Compounds* **2006**, *413*, 159.

[46] D. Di Lecce, V. Gancitano, J. Hassoun, *ACS Sustainable Chemistry and Engineering* **2020**, *8*, 278.

[47] M. Zhang, N. Garcia-Araez, A. L. Hector, *Journal of Materials Chemistry A* **2018**, *6*, 14483.

[48] W. Il Jung, M. Nagao, C. Pitteloud, A. Yamada, R. Kanno, *Journal of Power Sources* **2010**, *195*, 3328.

[49] J. Lehr, W. M. Dose, M. Yakovleva, S. W. Donne, *Journal of The Electrochemical Society* **2012**, *159*, A904.

[50] H. Tan, S. Wang, X. Lei, *Journal of The Electrochemical Society* **2015**, *162*, A448.

[51] V. Aravindan, J. Gnanaraj, S. Madhavi, H.-K. Liu, *Chemistry - A European Journal* **2011**, *17*, 14326.

[52] S. Lou, Y. Zhao, J. Wang, G. Yin, C. Du, X. Sun, *Small* **2019**, *15*, 1904740.

[53] B. Boukamp, *Solid State Ionics* **1986**, *20*, 31.

[54] B. A. Boukamp, *Solid State Ionics* **1986**, *18–19*, 136.

[55] S. Wei, D. Di Lecce, R. Messini D'Agostini, J. Hassoun, *ACS Applied Energy Materials* **2021**, *4*, 8340.

[56] X. Tao, J. Wang, C. Liu, H. Wang, H. Yao, G. Zheng, Z. W. Seh, Q. Cai, W. Li, G. Zhou, C. Zu, Y. Cui, *Nature Communications* **2016**, *7*, 11203.

[57] M. U. M. Patel, R. Demir-Cakan, M. Morcrette, J.-M. Tarascon, M. Gaberscek, R. Dominko, *ChemSusChem* **2013**, *6*, 1177.

[58] V. Marangon, J. Hassoun, *Energy Technology* **2019**, *7*, DOI 10.1002/ente.201900081.

[59] V. Marangon, D. Di Lecce, F. Orsatti, D. J. L. Brett, P. R. Shearing, J. Hassoun, *Sustainable Energy & Fuels* **2020**, *4*, 2907.

[60] V. Marangon, D. Di Lecce, D. J. L. Brett, P. R. Shearing, J. Hassoun, *Journal of Energy Chemistry* **2022**, *64*, 116.

[61] A. Benítez, P. Márquez, M. Á. Martín, A. Caballero, *ChemSusChem* **2021**, *14*, 3915.

[62] J. Xiao, J. Z. Hu, H. Chen, M. Vijayakumar, J. Zheng, H. Pan, E. D. Walter, M. Hu, X. Deng, J. Feng, B. Y. Liaw, M. Gu, Z. D. Deng, D. Lu, S. Xu, C. Wang, J. Liu, *Nano Letters* **2015**, *15*, 3309.

[63] C. fringant, A. Tranchant, R. Messina, *Electrochimica Acta* **1995**, *40*, 513.

[64] X. Tao, J. Wang, Z. Ying, Q. Cai, G. Zheng, Y. Gan, H. Huang, Y. Xia, C. Liang, W. Zhang, Y. Cui, *Nano Letters* **2014**, *14*, 5288.

[65] A. Bhargav, J. He, A. Gupta, A. Manthiram, *Joule* **2020**, *4*, 285.

[66] C. Cheng, S. Chung, *Batteries & Supercaps* **2022**, DOI 10.1002/batt.202100323.

[67] S.-H. Chung, C.-H. Chang, A. Manthiram, *Journal of Power Sources* **2016**, *334*, 179.

[68] R. Fang, S. Zhao, P. Hou, M. Cheng, S. Wang, H.-M. Cheng, C. Liu, F. Li, *Advanced Materials* **2016**, *28*, 3374.

[69] H. Wei, E. F. Rodriguez, A. S. Best, A. F. Hollenkamp, D. Chen, R. A. Caruso, *Advanced Energy Materials* **2017**, *7*, 1601616.

[70] X. Xu, S. Wang, H. Wang, B. Xu, C. Hu, Y. Jin, J. Liu, H. Yan, *Journal of Energy Storage* **2017**, *13*, 387.

[71] D. di Lecce, V. Marangon, A. Benítez, Á. Caballero, J. Morales, E. Rodríguez-Castellón, J. Hassoun, *Journal of Power Sources* **2019**, *412*, 575.

**List of tables**

**Table 1.** Non-linear least squares (NLLS) analyses performed through the Boukamp software on the Nyquist plots reported in Figure 1.[53,54]

**Table 2.** Non-linear least squares (NLLS) analyses performed through the Boukamp software on the Nyquist plots reported in Figure 2.[53,54]

**List of Figures**

**Figure 1.** Electrochemical behavior of **(a, c, e)** $MnO_2$ and **(b, d, f)** $TiO_2$ electrodes in lithium cell with the configuration Li|EC:DMC 1:1 V:V, 1M $LiPF_6$|cathode. In particular: **(a, b)** CV measurements performed at the scan rate of 0.1 mV $s^{-1}$ between 1.9 and 3.8 V *vs.* $Li^+$/Li for $MnO_2$ and from 1.4 to 2.8 V *vs.* $Li^+$/Li for $TiO_2$; **(c, d)** voltage profiles related to galvanostatic cycling of the lithium cells at the constant current rate of C/10 in the 1.9 – 3.8 V and 1.4 – 2.8 V voltage ranges for $MnO_2$ (1C = 308 mA $g^{-1}$) and $TiO_2$ (1C = 168 mA $g^{-1}$), respectively; **(e, f)** Nyquist plots recorded through EIS carried out on the lithium cells at the OCV condition and upon 10 discharge/charge cycles. Frequency range: 500 kHz – 10 mHz. Alternate voltage signal amplitude: 10 mV. Room temperature (25 °C)

**Figure 2. (a)** Samples of the DOL:DME, 1 mol $kg^{-1}$ LiTFSI, 1 mol $kg^{-1}$ $LiNO_3$ electrolyte solution added with the 0.5 wt.% of $Li_2S_8$ (see Experimental section for preparation details) either at the pristine state (left-hand side), or upon 5 hours aging in contact with $MnO_2$ (center) and $TiO_2$ (right-hand side), additional images are reported in Fig. S3 (Supporting Information); **(b)** UV-vis

measurements performed on the $Li_2S_8$-containing solutions displayed in panel **(a)** in the 500 – 800 nm range by using a DOL.DME 1:1 *w/w* solution as reference; **(c, d)** XRD patterns of the **(c)** S-$MnO_2$ and **(d)** S-$TiO_2$ sulfur-metal oxide composites, reference data for orthorhombic sulfur ($S_8$, ICSD #27840, yellow), manganese oxide ($\beta$-$MnO_2$, ICSD #73716, violet) and titanium oxide ($\alpha$-$TiO_2$, ICSD #154607, blue) are also reported; **(e, f)** TGA and corresponding DTG carried out on **(e)** the S-$MnO_2$ and **(f)** S-$TiO_2$ composites between 25 and 800 °C under $N_2$ flow at an increasing temperature of 5 °C $min^{-1}$. See experimental section for acronyms.

**Figure 3.** SEM images of the **(a, b)** S-$MnO_2$ and **(f, g**) S-$TiO_2$ composites with corresponding EDS elemental maps in panels **(c-e)** and **(h-j)**, respectively, showing distribution of **(c)** manganese, **(h)** titanium, **(d, i)** oxygen, and **(e, j)** sulfur. See experimental section for acronyms.

**Figure 4.** Electrochemical behavior of the **(a, c)** S-$MnO_2$ and **(b, d)** S-$TiO_2$ composites in lithium cell with the configuration Li|DOL:DME, 1 mol $kg^{-1}$ LiTFSI, 1 mol $kg^{-1}$ $LiNO_3$|cathode. In particular: **(a, b)** CV tests performed at the scan rate of 0.1 mV $s^{-1}$ in the 1.8 – 2.8 V *vs.* $Li^+$/Li potential range; **(c, d)** Nyquist plots recorded through EIS carried out on the lithium cells at the OCV condition (inset) and after 1, 5 and 10 CV cycles. Frequency range: 500 kHz – 100 mHz. Alternate voltage signal amplitude: 10 mV. Sulfur loading: ~1.5 mg $cm^{-2}$ (electrode geometric area: 1.54 $cm^{-2}$). Electrolyte/sulfur (E/S) ratio: 15 µL $mg^{-1}$. Room temperature (25 °C). See experimental section for acronyms.

**Figure 5. (a, b)** Voltage profiles and **(c, d)** corresponding cycling trends collected upon rate capability tests of lithium cells with the configuration Li|DOL:DME, 1 mol $kg^{-1}$ LiTFSI, 1 mol $kg^{-1}$ $LiNO_3$|cathode, employing either the **(a, c)** S-$MnO_2$ or **(b, d)** S-$TiO_2$ electrode. C-rates: C/10, C/8, C/5, C/3, C/2, 1C and 2C. Voltage ranges: 1.9 – 2.8 V from C/10 to C/2 and 1.8 – 2.8 V for 1C and 2C (1C = 1675 mA $g_S^{-1}$). Sulfur loading: 2 mg $cm^{-2}$ (electrode geometric area: 1.54 $cm^{-2}$). Electrolyte/sulfur (E/S) ratio: 15 µL $mg^{-1}$. Room temperature (25 °C). See experimental section for acronyms.

**Figure 6.** Trends of the prolonged galvanostatic cycling tests performed on lithium cells with the configuration Li|DOL:DME, 1 mol $kg^{-1}$ LiTFSI, 1 mol $kg^{-1}$ $LiNO_3$|cathode, employing either S-$MnO_2$ (red) or S-$TiO_2$ (green) electrodes at the current rates of **(a)** C/5, **(b)** 1C and **(c)** 2C (the latter only for S-$TiO_2$). Voltage ranges: 1.9 – 2.8 V for C/5 and 1.8 – 2.8 V for 1C and 2C. Sulfur loading: 1.5 – 2 mg $cm^{-2}$ (electrode geometric area: 1.54 $cm^{-2}$). Electrolyte/sulfur (E/S) ratio: 15 µL $mg^{-1}$. The corresponding voltage profiles are reported in Figure S4 in the Supporting Information. Room temperature (25 °C). See experimental section for acronyms.

**Figure 7.** Galvanostatic cycling tests of lithium cells with the configuration Li|DOL:DME, 1 mol $kg^{-1}$ LiTFSI, 1 mol $kg^{-1}$ $LiNO_3$|cathode, employing either the **(a, c)** S-$MnO_2$ or **(b, d)** S-$TiO_2$ electrode with the relatively high sulfur loading of 5.4 and 5.8 mg $cm^{-2}$, respectively (electrode geometric area: 1.54 $cm^{-2}$). In particular: **(a, b)** voltage profiles and **(c, d)** corresponding cycling trend of discharge/charge cycling at the constant current rate of C/5 (1C = 1675 mA $g_S^{-1}$) in the 1.7 – 2.8 V voltage range. Inset reports corresponding Coulombic Efficiencies. Electrolyte/sulfur (E/S) ratio: 10 µL $mg^{-1}$. Room temperature (25 °C). See experimental section for acronyms.

| Material | Cell condition | Circuit | $R_i$ [Ω] | $\chi^2$ |
|---|---|---|---|---|
| $MnO_2$ | OCV | $R_e(R_iQ_i)Q_g$ | 186 ± 1 | $3\times10^{-4}$ |
| | After 10 cycles | $R_e(R_iQ_i)Q_g$ | 131 ± 1 | $5\times10^{-4}$ |
| $TiO_2$ | OCV | $R_e(R_iQ_i)Q_g$ | 77.2 ± 0.3 | $9\times10^{-5}$ |
| | After 10 cycles | $R_e(R_iQ_i)Q_g$ | 37.1 ± 0.4 | $7\times10^{-4}$ |

**Table 1**

| Material | Cell condition | Equivalent Circuit | $R_1$ [Ω] | $R_2$ [Ω] | $R_3$ [Ω] | $R_i$ ($\sum R_n$) [Ω] | $\chi^2$ |
|---|---|---|---|---|---|---|---|
| $S:MnO_2$ 80:20 w/w | OCV | $R_e(R_1Q_1)(R_2Q_2)Q_g$ | 22.5 ± 0.1 | 1.9 ± 0.4 | / | 24.4 ± 0.5 | $5\times10^{-5}$ |
| | After 1 CV cycle | $R_e(R_1Q_1)(R_2Q_2)Q_g$ | 5.1 ± 0.3 | 1.8 ± 0.5 | / | 6.9 ± 0.8 | $8\times10^{-5}$ |
| | After 5 CV cycles | $R_e(R_1Q_1)(R_2Q_2)Q_g$ | 5.8 ± 0.2 | 0.4 ± 0.1 | / | 6.2 ± 0.3 | $3\times10^{-5}$ |
| | After 10 CV cycles | $R_e(R_1Q_1)(R_2Q_2)(R_3Q_3)Q_g$ | 1.4 ± 0.2 | 3.9 ± 0.3 | 0.6 ± 0.1 | 5.9 ± 0.6 | $1\times10^{-5}$ |
| $S:TiO_2$ 80:20 w/w | OCV | $R_e(R_1Q_1)(R_2Q_2)$ | 9.9 ± 0.2 | 4.2 ± 0.5 | / | 14.1 ± 0.7 | $5\times10^{-5}$ |
| | After 1 CV cycle | $R_e(R_1Q_1)(R_2Q_2)Q_w$ | 2.0 ± 0.1 | 1.5 ± 0.5 | / | 3.5 ± 0.6 | $9\times10^{-5}$ |
| | After 5 CV cycles | $R_e(R_1Q_1)(R_2Q_2)Q_w$ | 2.6 ± 0.1 | 0.9 ± 0.2 | / | 3.5 ± 0.3 | $2\times10^{-5}$ |
| | After 10 CV cycles | $R_e(R_1Q_1)(R_2Q_2)(R_3Q_3)Q_w$ | 0.6 ± 0.2 | 4.2 ± 0.2 | 0.5 ± 0.1 | 5.3 ± 0.5 | $5\times10^{-5}$ |

**Table 2**

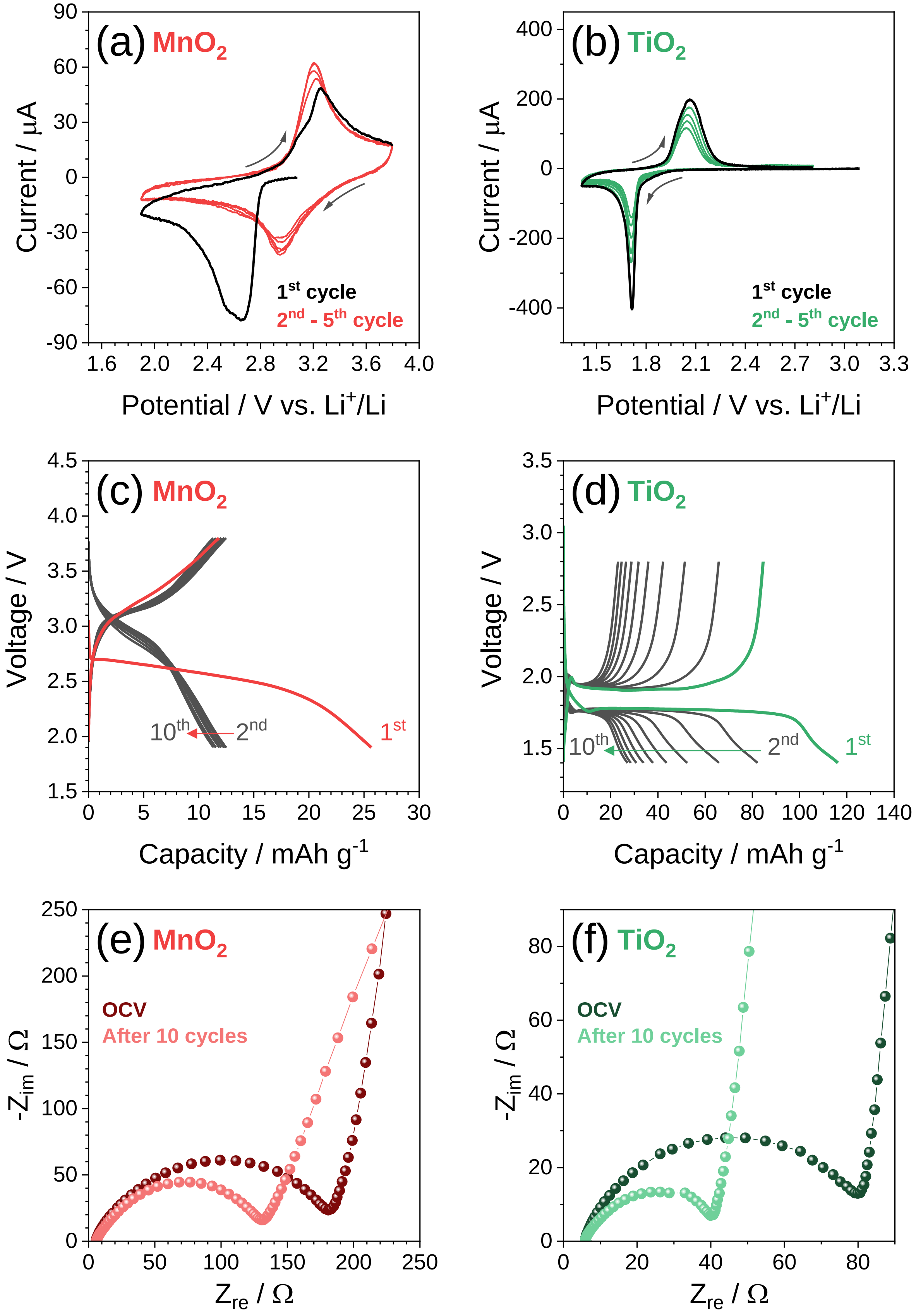

**Figure 1**

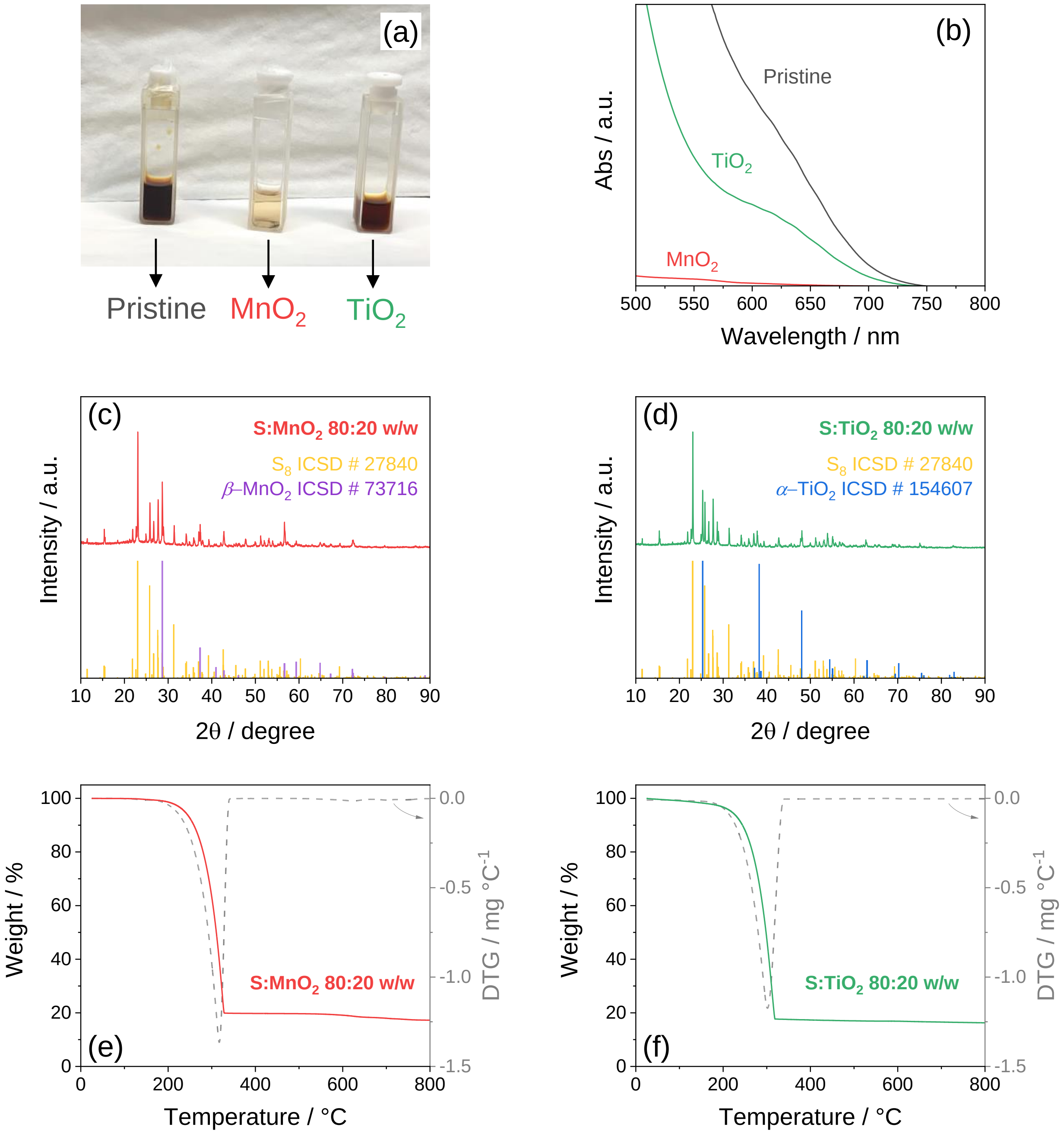
(a)
Pristine
MnO2
TiO2
(b)
Abs / a.u.
Pristine
TiO2
MnO2
500 550 600 650 700 750 800
Wavelength / nm
(c)
S:MnO2 80:20 w/w
S8 ICSD # 27840
β–MnO2 ICSD # 73716
Intensity / a.u.
10 20 30 40 50 60 70 80 90
2θ / degree
(d)
S:TiO2 80:20 w/w
S8 ICSD # 27840
α–TiO2 ICSD # 154607
Intensity / a.u.
10 20 30 40 50 60 70 80 90
2θ / degree
(e)
S:MnO2 80:20 w/w
Weight / %
DTG / mg °C-1
Temperature / °C
(f)
S:TiO2 80:20 w/w
Weight / %
DTG / mg °C-1
Temperature / °C

**Figure 2**

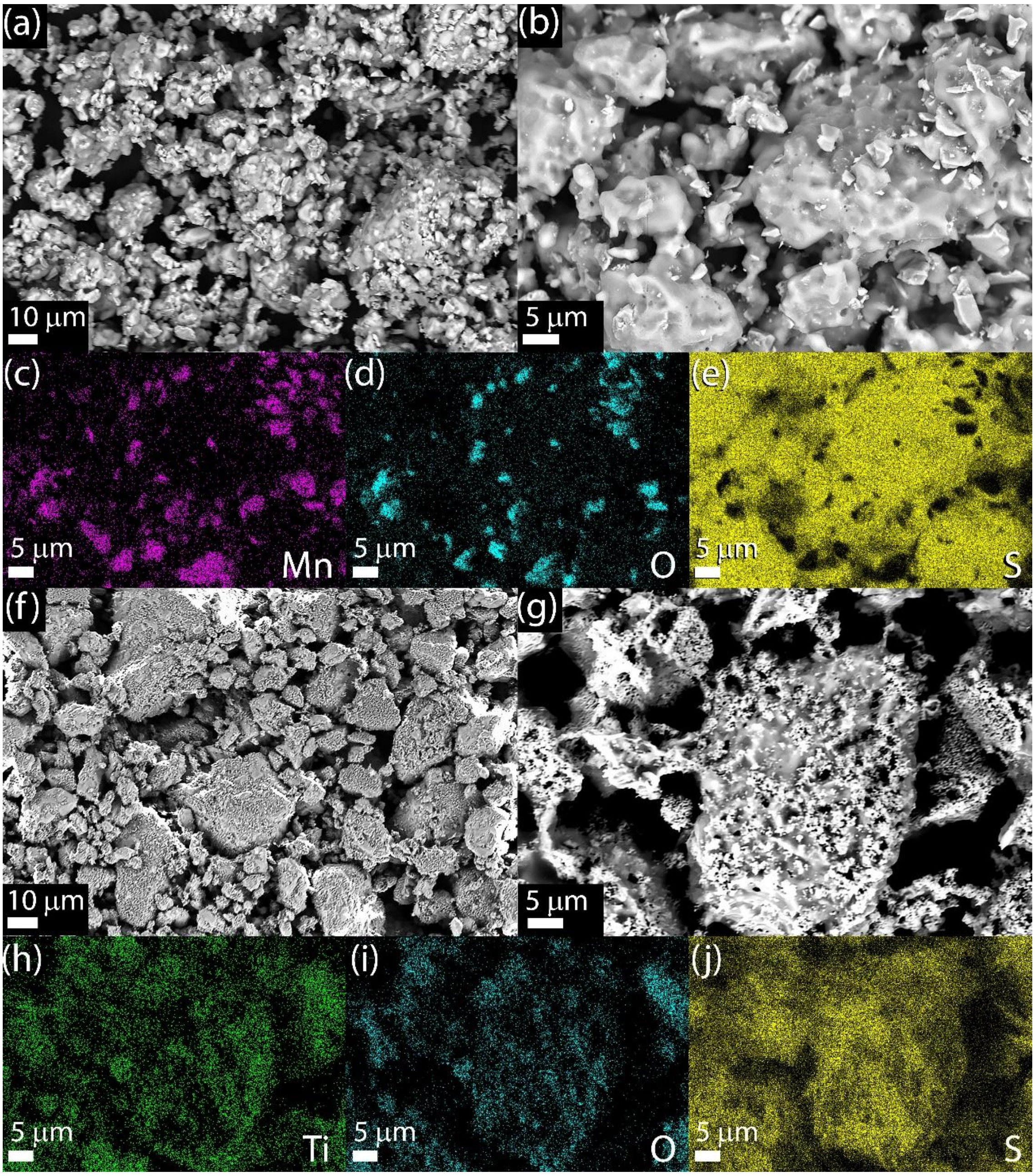

**Figure 3**

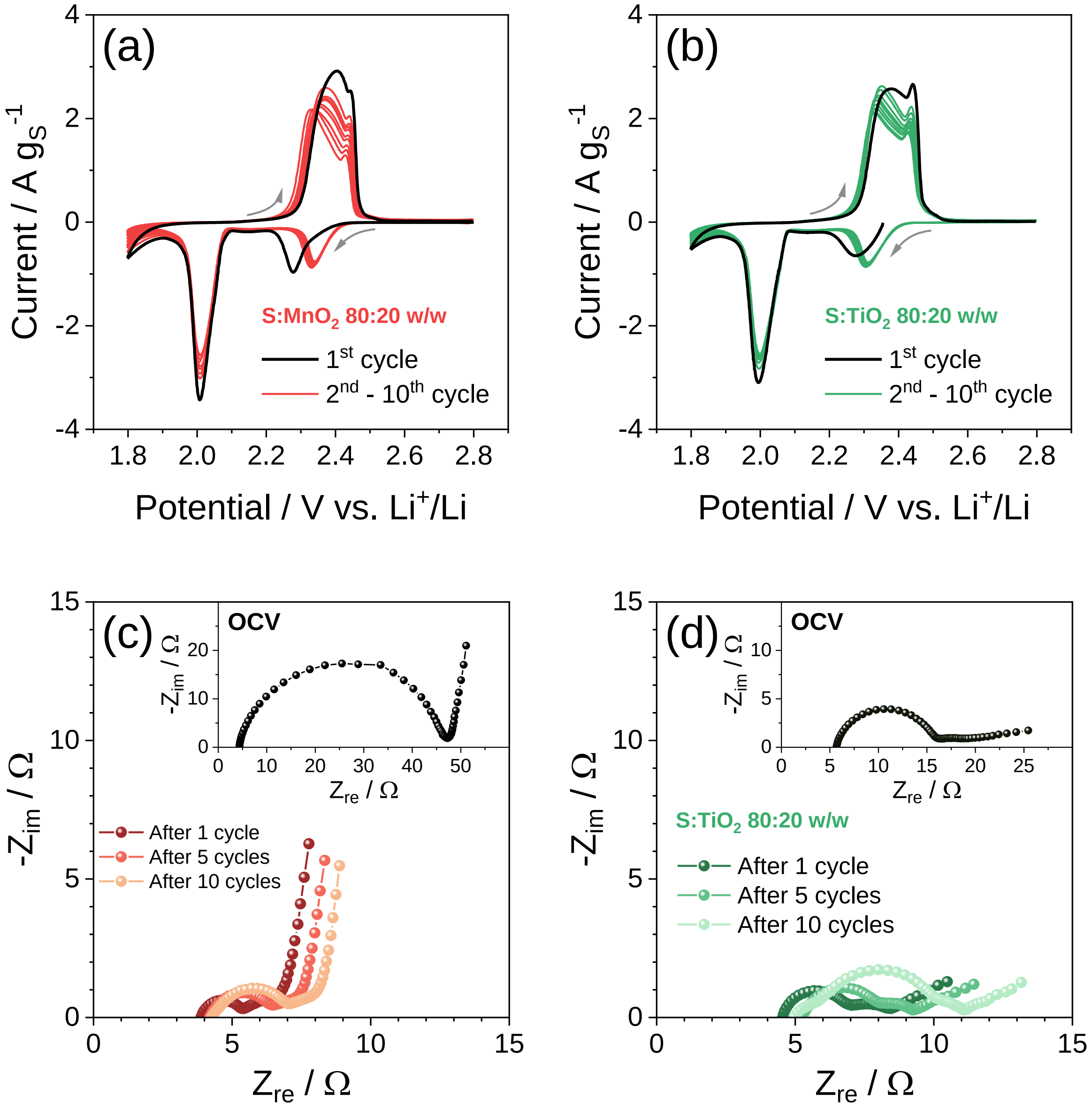
(a)
Current / A $g_S^{-1}$
Potential / V vs. $Li^+$/Li
S:$MnO_2$ 80:20 w/w
$1^{st}$ cycle
$2^{nd}$ - $10^{th}$ cycle
(b)
Current / A $g_S^{-1}$
Potential / V vs. $Li^+$/Li
S:$TiO_2$ 80:20 w/w
$1^{st}$ cycle
$2^{nd}$ - $10^{th}$ cycle
(c)
OCV
$-Z_{im}$ / Ω
$Z_{re}$ / Ω
After 1 cycle
After 5 cycles
After 10 cycles
(d)
OCV
$-Z_{im}$ / Ω
$Z_{re}$ / Ω
S:$TiO_2$ 80:20 w/w
After 1 cycle
After 5 cycles
After 10 cycles

**Figure 4**

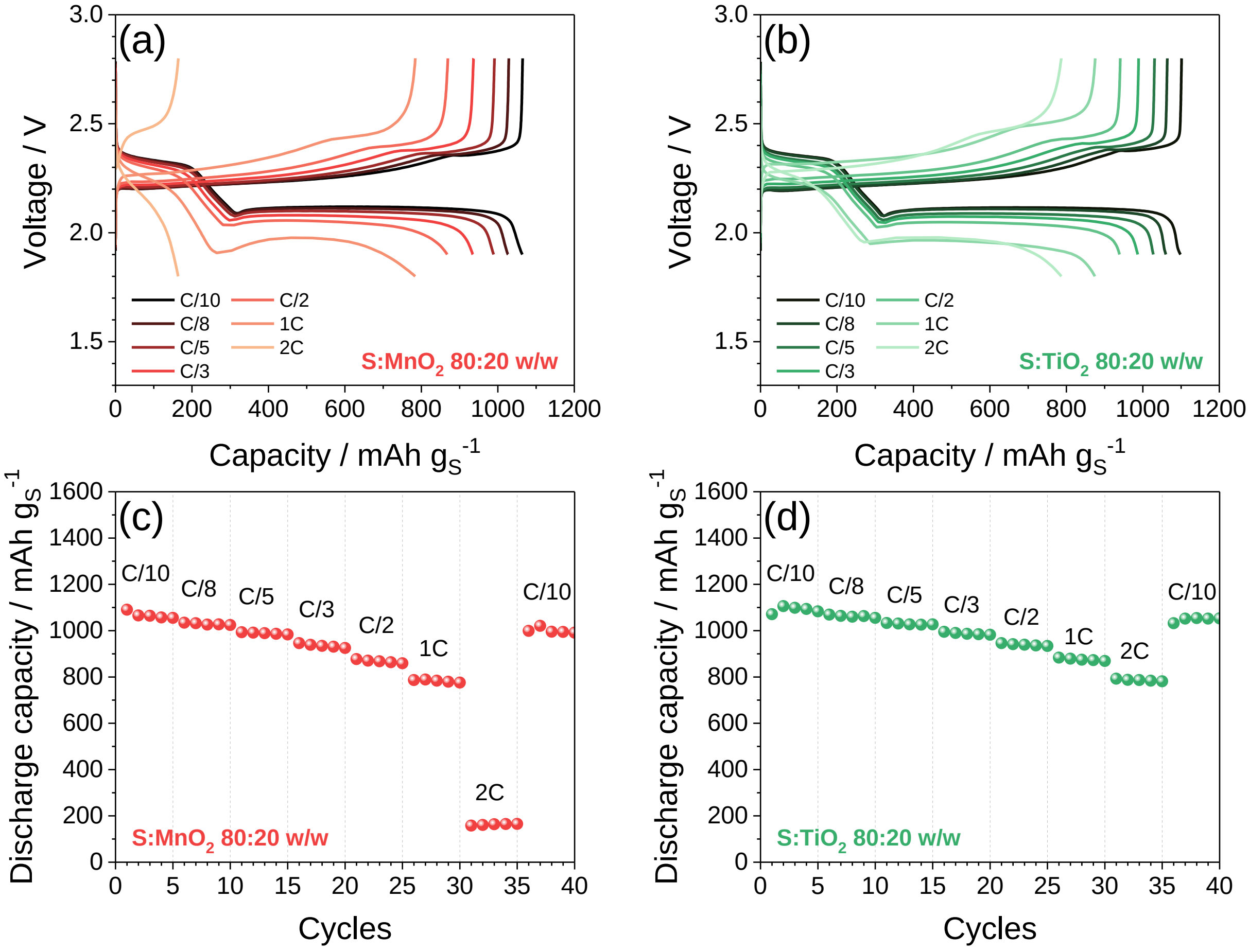

**Figure 5**

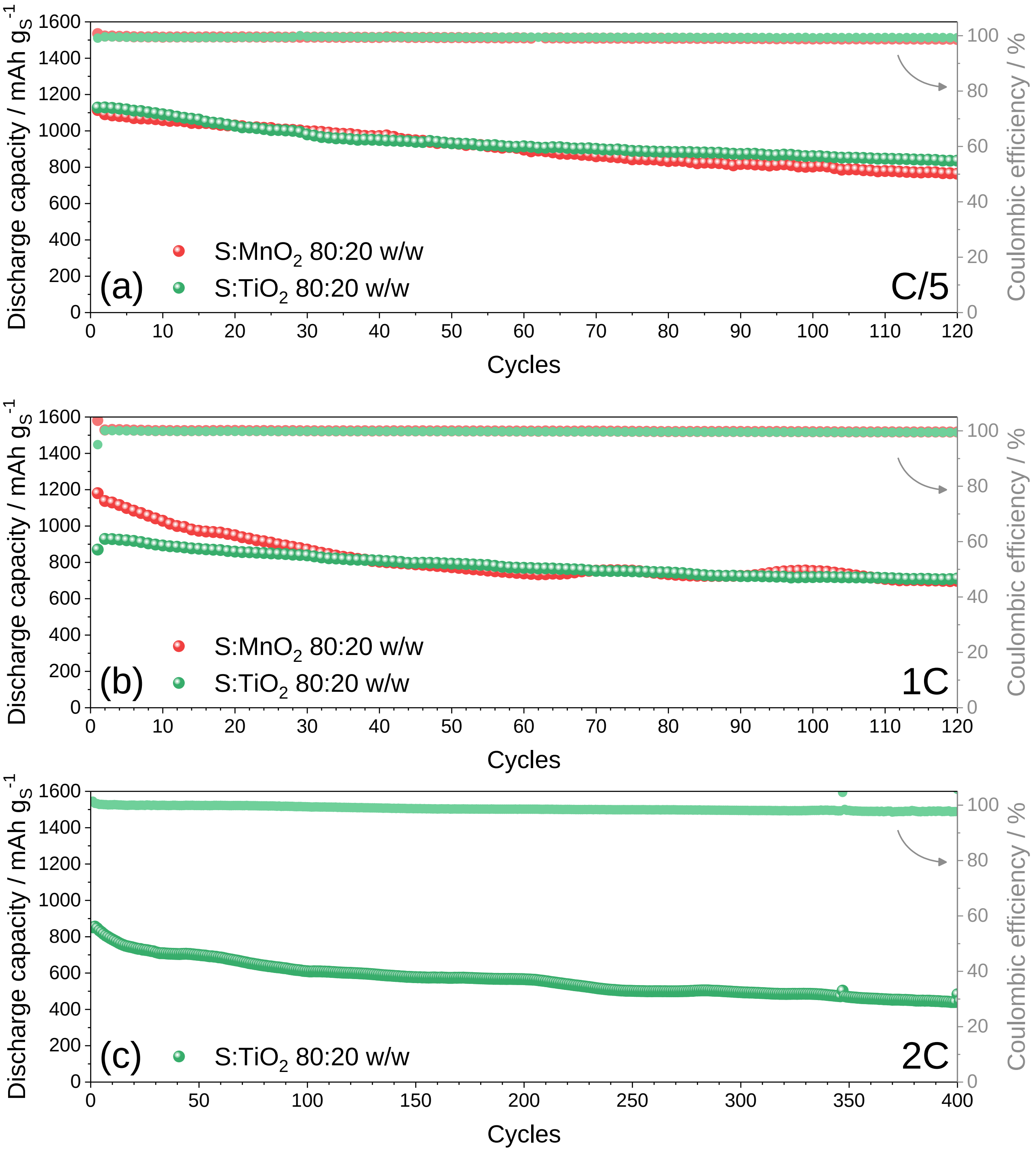

**Figure 6**

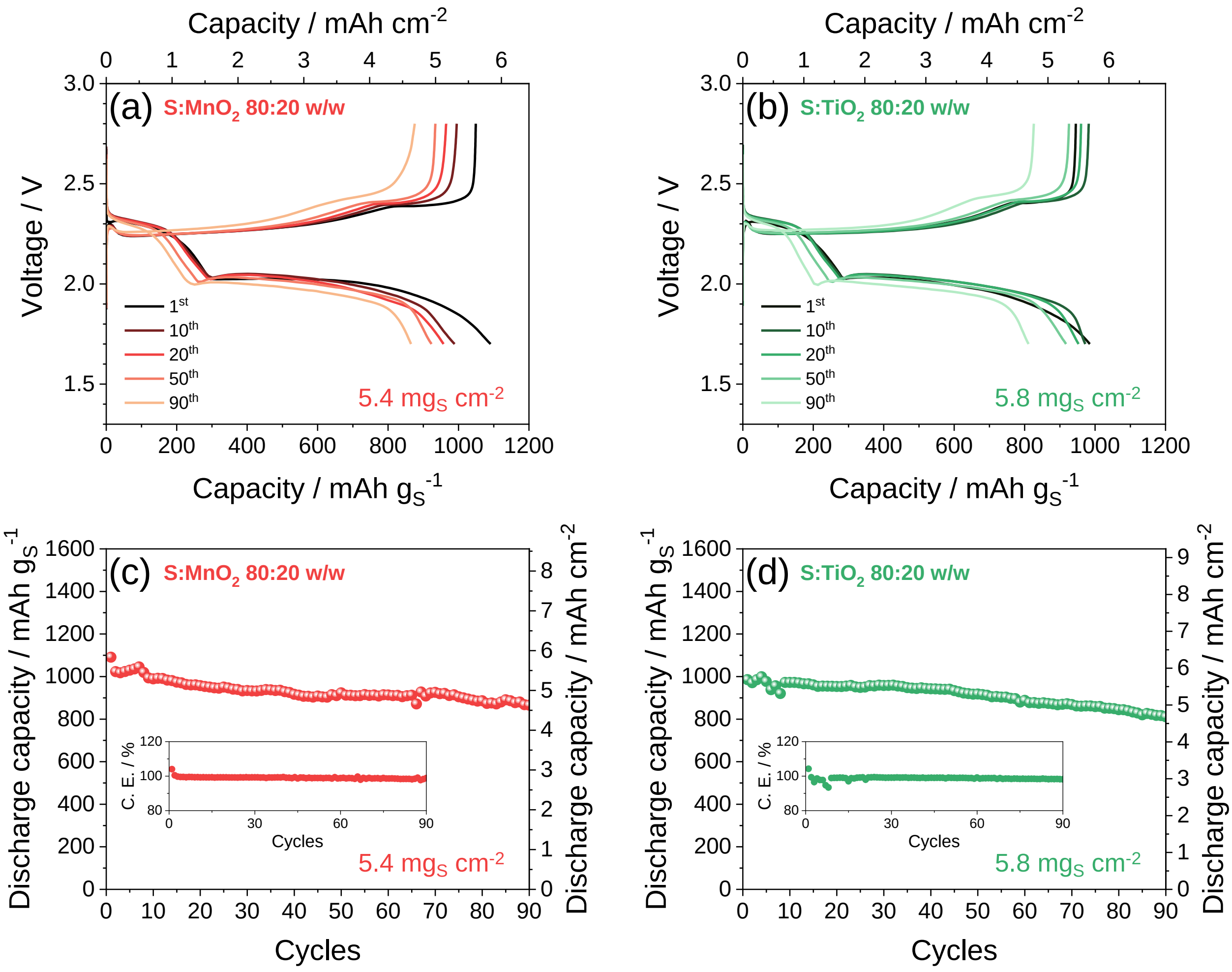

Capacity / mAh cm-2
(a) S:MnO2 80:20 w/w
(b) S:TiO2 80:20 w/w
Voltage / V
1st
10th
20th
50th
90th
5.4 mgS cm-2
5.8 mgS cm-2
Capacity / mAh gS-1
(c) S:MnO2 80:20 w/w
(d) S:TiO2 80:20 w/w
Discharge capacity / mAh gS-1
Discharge capacity / mAh cm-2
C. E. / %
Cycles

**Figure 7**

Figure S1 shows the XRD (Fig. S1a, b) and SEM (Fig. S1c-h) investigation of the $MnO_2$ and $TiO_2$ powders. The diffractograms identify for $MnO_2$ the $\beta$ crystalline phase, that is, pyrolusite (Fig. S1a) with $P4_2/mnm$ space group,[1] and the $\alpha$ one, that is, anatase, for $TiO_2$ (Fig. S1b) with $I4_1/amdZ$ space group.[2] The SEM images reveal substantial differences between the metal oxide morphologies, and show micrometric aggregates with defined submicron flake-like geometry for $MnO_2$ (Fig. S1c, e and g) and nano-sized particles for $TiO_2$ (Fig. S1d, f and h). The micrometric morphology may hinder electrolyte degradation, while nanometric particles may shorten the charge-transfer path and enhance the rection kinetics in lithium cell.

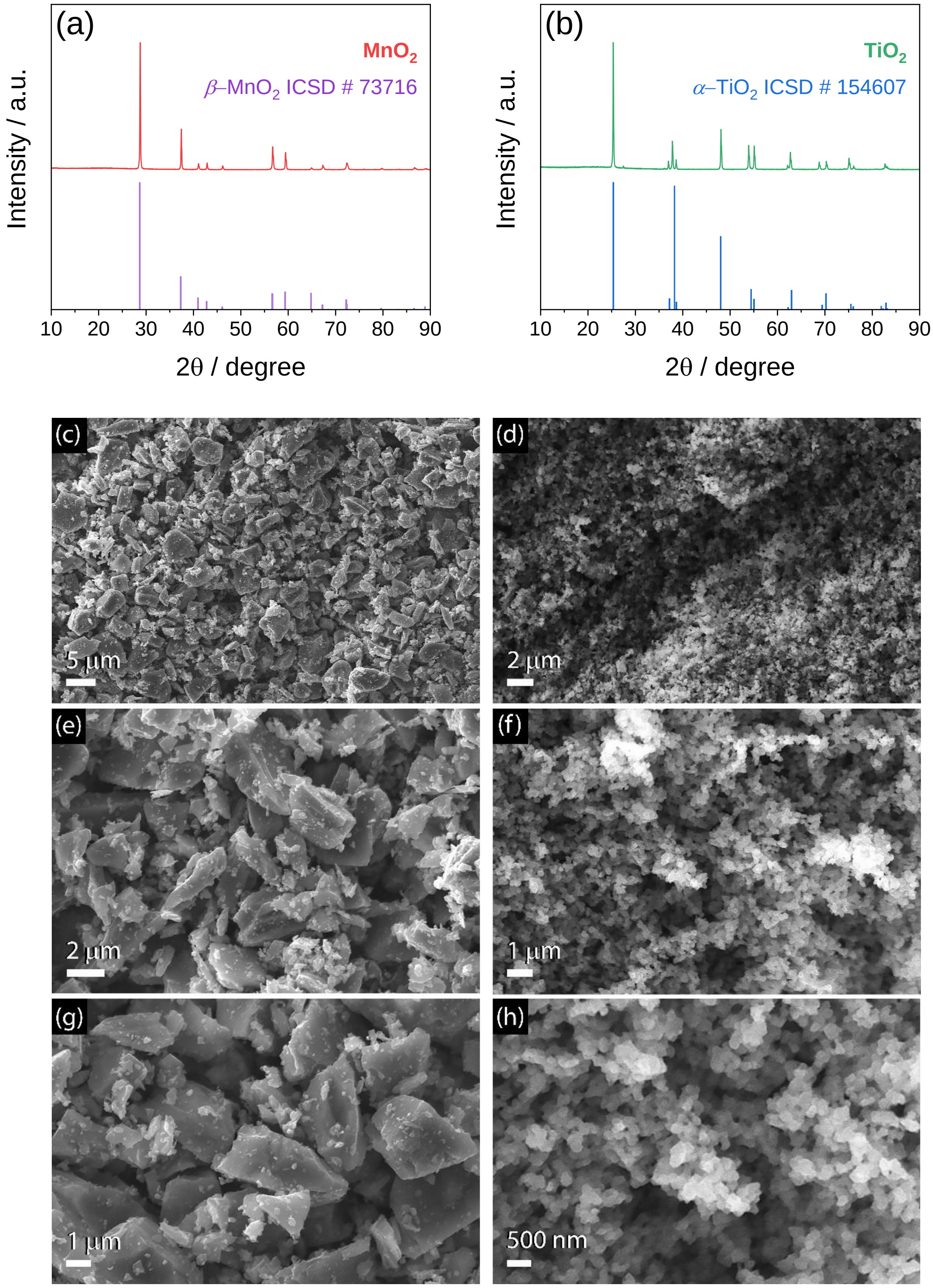

**Figure S1. (a, b)** XRD patterns of **(a)** $MnO_2$ and **(b)** $TiO_2$ samples. The reference data for $\beta$-$MnO_2$ (pyrolusite, ICSD # 73716, violet) and $\alpha$-$TiO_2$ (anatase, ICSD # 154607, blue) are reported for comparison. **(c-h)** SEM images of **(c, e, g)** $MnO_2$ and **(d, f, h)** $TiO_2$ powders at various magnifications.

TEM analyses of Figure S2 confirm the vast size range of the $MnO_2$ flakes, extending from 50 nm to 3 µm, and show wide $TiO_2$ clusters composed of quasi-spherical primary particles with dimensions from 50 to 200 nm.

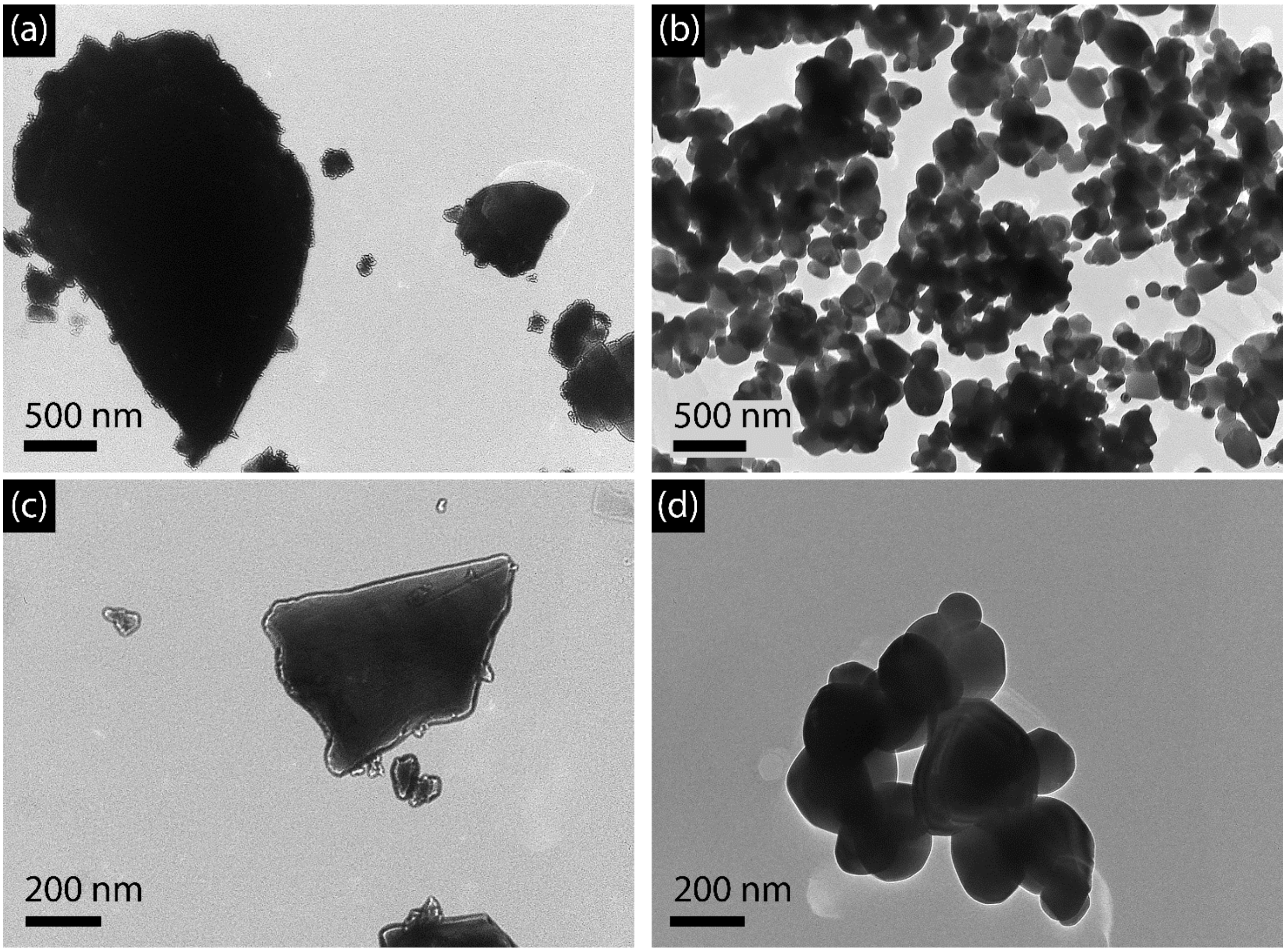

**Figure S2.** TEM images at various magnifications of **(a, c)** $MnO_2$ and **(b, d)** $TiO_2$ powders.

The lithium polysulfides retention ability of the $MnO_2$ and $TiO_2$ powders is qualitatively evaluated in Figure S3. Vials containing $MnO_2$ and $TiO_2$ (Fig. S3a) are filled with the DOL:DME, 1 mol $kg^{-1}$ LiTFSI, 1 mol $kg^{-1}$ $LiNO_3$, 0.5 wt.% $Li_2S_8$ electrolyte and the resulting solutions are compared with the pristine one after 90 minutes aging (Fig. S3b). The solutions held in contact with $MnO_2$ and $TiO_2$ show a different color compared to the pristine one, i.e., light yellow and bright red rather than dark red, respectively. These variations are due to the more effective polysulfides retention of $MnO_2$ with respect to $TiO_2$, which is likely related to the metal oxide nature and to the oxide morphology.

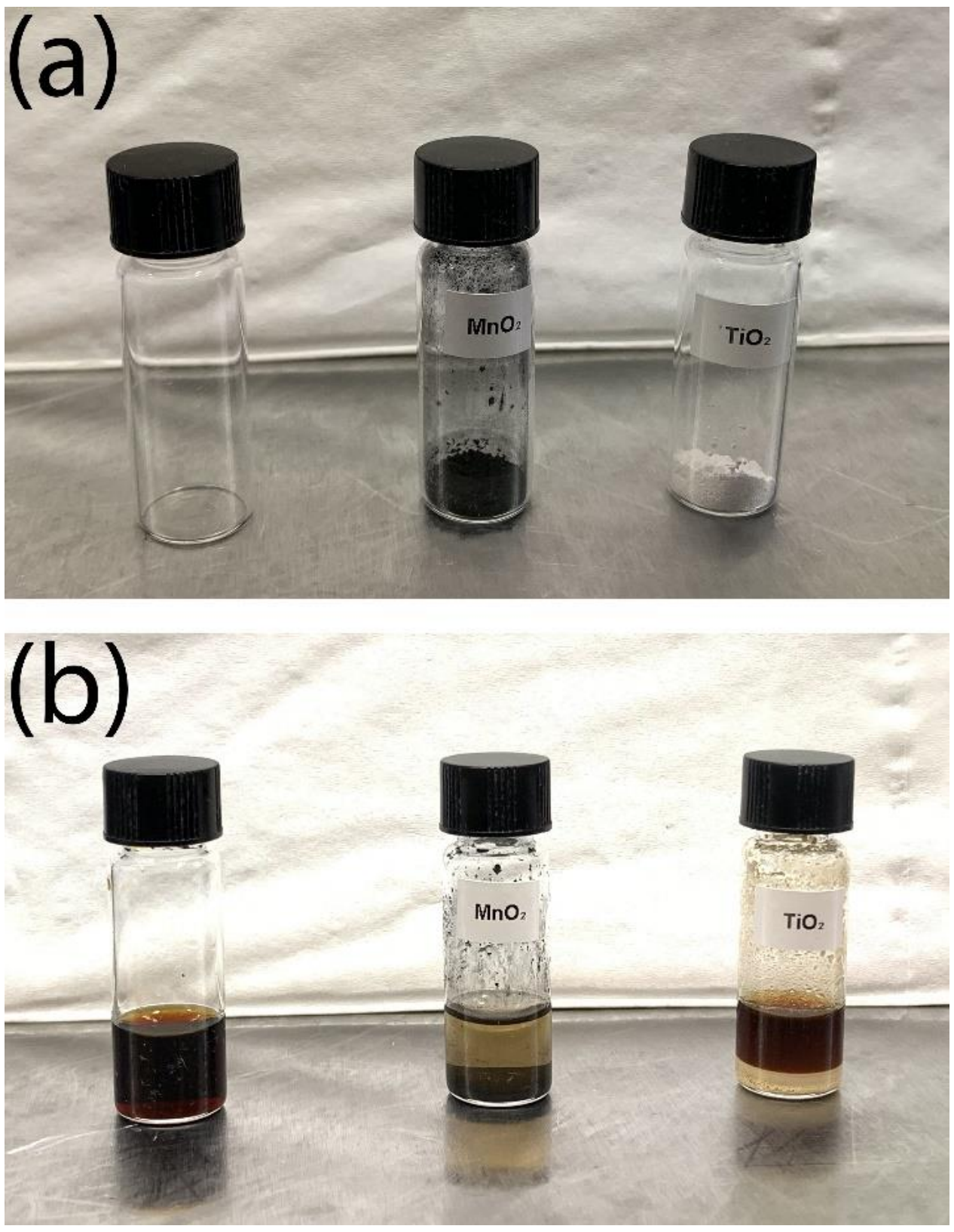

**Figure S3.** Comparison of the polysulfides retention ability between $MnO_2$ and $TiO_2$ by meaning of photographic images. Panel **(a)** shows empty bottle (left-hand side), and bottles containing $MnO_2$ (center) and hosting $TiO_2$ (right-hand side). Panel **(b)** displays the same bottles filled by the $Li_2S_8$_added electrolyte solution upon 90 minutes of aging (see experimental section in the manuscript for details on composition).

The differences observed between the transition metal oxides are reflected into the electrochemical responses in lithium cell of the S-$MnO_2$ and S-$TiO_2$ sulfur composites synthetized by straightforward physical mixing of elemental sulfur with either $MnO_2$ or $TiO_2$ (see Experimental section for details), as displayed in Figure S4. The voltage profiles corresponding to the galvanostatic cycling tests performed on Li|S-$MnO_2$ (Fig. S4a, c and e) and Li|S-$TiO_2$ (Fig. S4b, d and f) cells evidence similar responses of the at C/5 (1C = 1675 mA $g_S^{-1}$) with initial capacity values exceeding 1100 mAh $g_S^{-1}$ retained around the 70 % after 120 cycles (Fig. S4a, b). At the current of 1C, S-$MnO_2$ delivers a higher initial capacity with respect to S-$TiO_2$, which in turns, shows a better capacity retention upon 120 cycles, that is, 76 % (Fig. S4d) with respect to the 61 % of S-$MnO_2$ (Fig. S4c). This aspect is more evident when the C-rate is increased to 2C: indeed, the Li-S conversion process of the Li|S-$MnO_2$ cell takes place with a reversible capacity as low as 200 mAh $g_S^{-1}$ (Fig. S4e), while the Li|S-$TiO_2$ system delivers over 800 mAh $g^{-1}$ and exhibits 400 cycles and with a retention over 50 % at the end of the test (Fig. S4f). The corresponding cycling trends are reported in Figure 5 in the Manuscript. The notable electrochemical activity of the Li|S-$TiO_2$ cell at relatively high current rate with respect to the Li|S-$MnO_2$ one is likely attributed to a more relevant effect of the nanometric $TiO_2$ particles in boosting the conversion kinetics compared to the micrometric $MnO_2$.[3] Moreover, an excessive polysulfide retention during battery operation of the $MnO_2$ additive (see Fig. S3b) may possibly hinder the Li-S conversion kinetics and lower rate capability.[4]

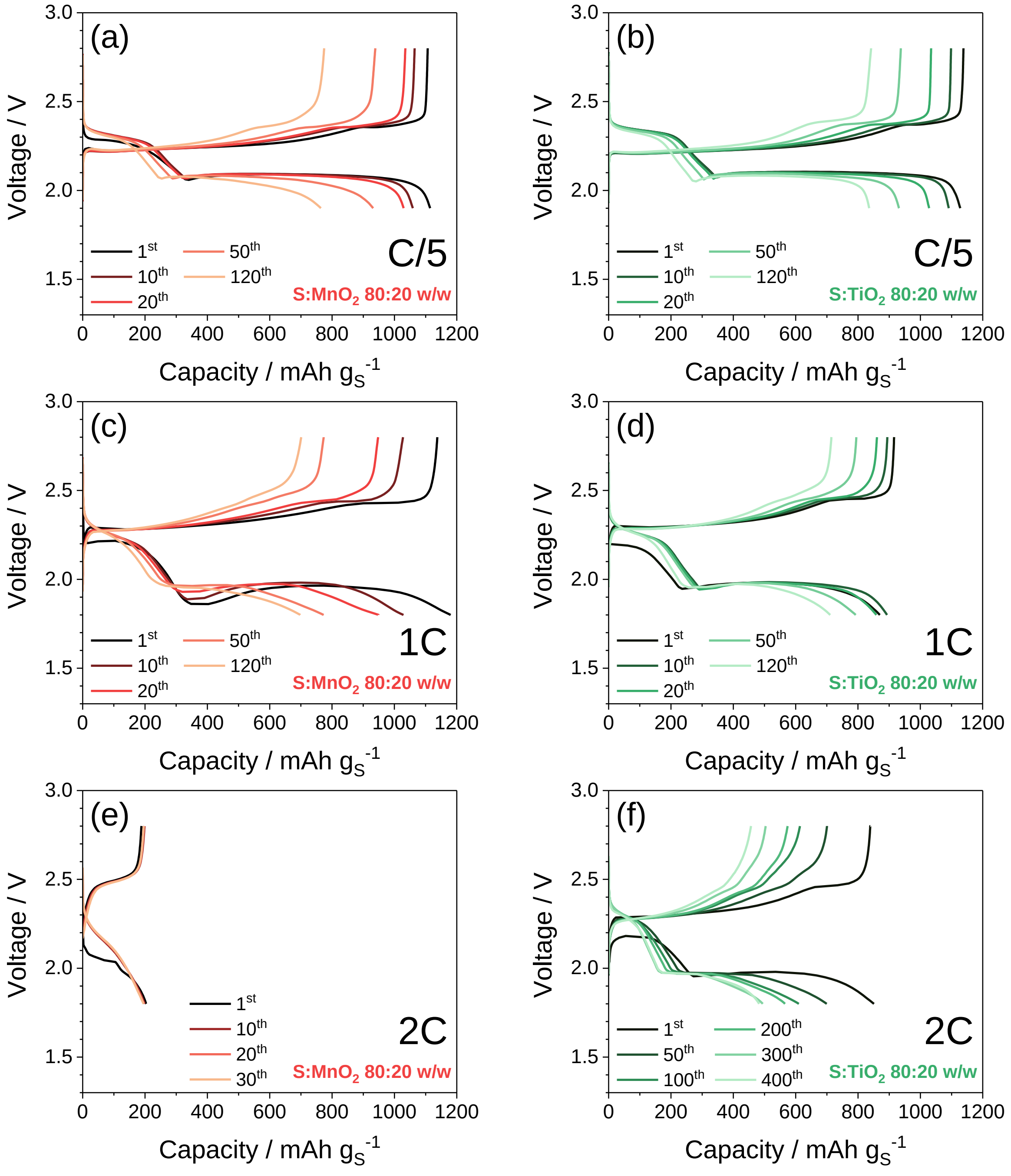

**Figure S4.** Voltage profiles related to galvanostatic cycling tests performed at the current rates of **(a, b)** C/5, **(c, d)** 1C and **(e, f)** 2C in Li|DOL:DME, 1 mol $kg^{-1}$ LiTFSI, 1 mol $kg^{-1}$ $LiNO_3$|cathode cells employing either **(a, c, e)** S-$MnO_2$ or **(b, d, f)** S-$TiO_2$ electrodes. Voltage ranges: 1.9 – 2.8 V for C/5 and 1.8 – 2.8 V for 1C and 2C. Sulfur loading: 1.5 – 2 mg $cm^{-2}$ (referred to the electrode geometric area of 1.54 $cm^{-2}$). Electrolyte to sulfur (E/S) ratio: 15 μL $mg^{-1}$. The corresponding cycling trends are reported in Figure 6 in the Manuscript. Room temperature (25 °C). See experimental section for acronyms.

Figure S5 reports the performance of Li cells employing a control sulfur cathode without metal oxides and with loading of either 5.0 (Fig. S5a, c) or 4.5 mg cm$^{-2}$ (Fig. S5b, d) and no $MnO_2$ or $TiO_2$ (S blank). Despite the acceptable initial cycling behavior, both the cells show several short circuits likely caused by dendritic structures formed on the Li surface. The massive diffusion of lithium polysulfides in absence of the restraining action of the metal oxide and their direct rection at the lithium surface cause the formation of an irregular SEI and metallic dendrites.

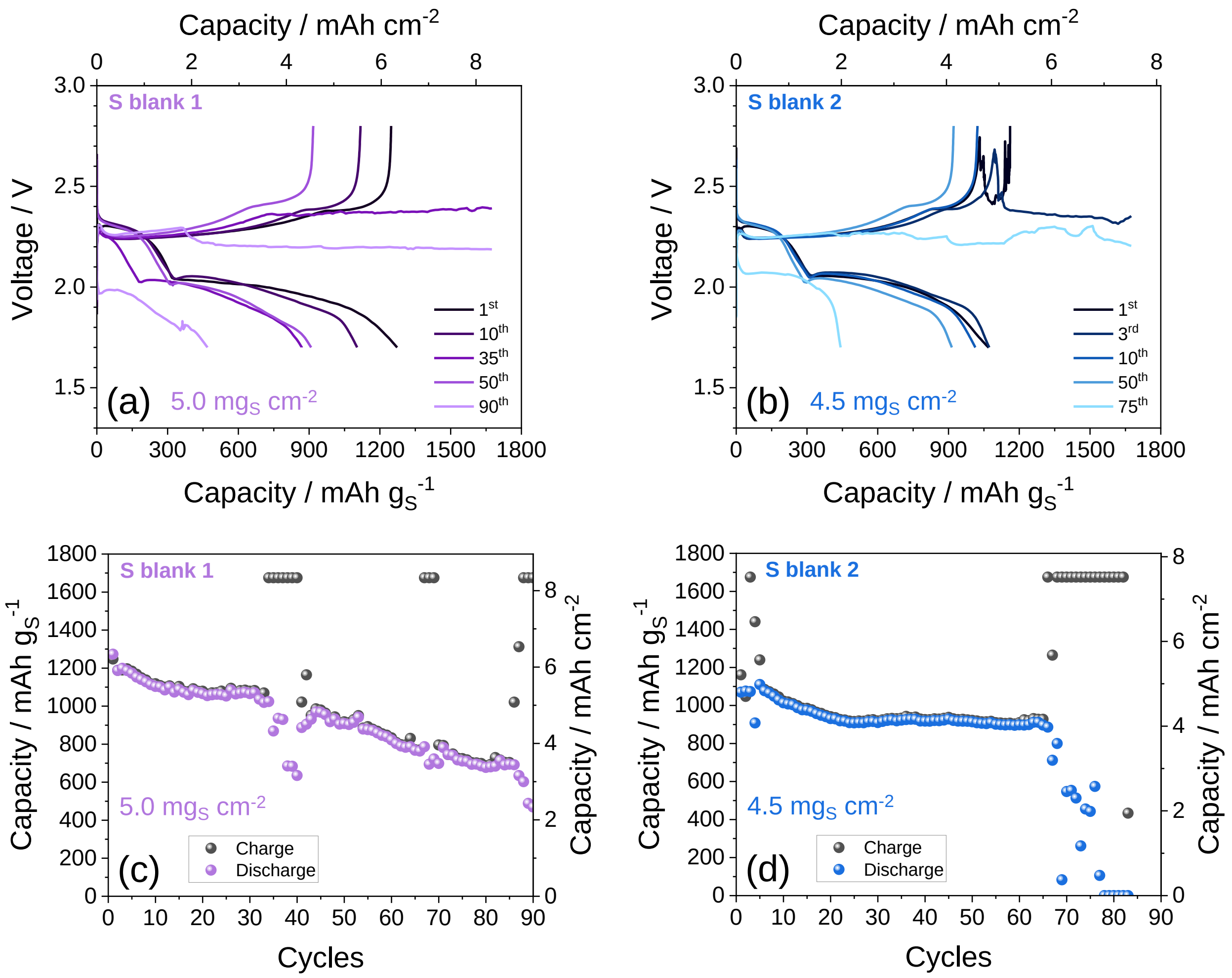

**Figure S5. (a, b)** Galvanostatic profiles and **(c, d)** corresponding galvanostatic trends of Li|DOL:DME, 1 mol kg$^{-1}$ LiTFSI, 1 mol kg$^{-1}$ $LiNO_3$|cathode employing a sulfur control electrode (see Experimental section) with a sulfur loading of either **(a, c)** 5.0 or **(b, d)** 4.5 $mg_S$ cm$^{-2}$ (electrode geometric area: 1.54 cm$^{-2}$) cycled at the constant current rate of C/5 in the 1.7 – 2.8 V voltage range. Electrolyte/sulfur (E/S) ratio: 10 µL mg$^{-1}$. Room temperature (25 °C). See experimental section for acronyms.

**References**

[1] A. Bolzan, C. Fong, B. Kennedy, C. Howard, *Australian Journal of Chemistry* **1993**, *46*, 939.

[2] I. Djerdj, A. M. Tonejc, *Journal of Alloys and Compounds* **2006**, *413*, 159.

[3] D. Di Lecce, V. Gancitano, J. Hassoun, *ACS Sustainable Chemistry and Engineering* **2020**, *8*, 278.

[4] D. di Lecce, V. Marangon, A. Benítez, Á. Caballero, J. Morales, E. Rodríguez-Castellón, J. Hassoun, *Journal of Power Sources* **2019**, *412*, 575.